\documentclass[aps,prc,showpacs,preprint]{revtex4}
\usepackage{graphicx}
\usepackage[pdftex,colorlinks,citecolor=blue,bookmarks]{hyperref}

\begin{document}
\title{Triaxial Angular Momentum Projection and Configuration Mixing calculations with the Gogny force}

\author{Tom\'as R. Rodr\'iguez}
\affiliation{GSI Helmholtzzentrum f\"ur Schwerionenforschung\footnote{Present address}, D-64259 Darmstadt, Germany}
\affiliation{Departamento de F\'isica Te\'orica, Universidad Aut\'onoma de Madrid,
E-28049 Madrid, Spain}
\author{J. Luis Egido}
\affiliation{Departamento de F\'isica Te\'orica, Universidad Aut\'onoma de Madrid,
E-28049 Madrid, Spain}
\date{\today}
\begin{abstract}
We present the first implementation in the $(\beta,\gamma)$ plane of the generator coordinate method with full triaxial angular momentum and particle number projected wave functions  using the Gogny force. Technical details about the performance of the method and the convergence of the results both in the symmetry restoration and the configuration mixing parts are discussed in detail.  We apply the method to the study of $^{24}$Mg, the calculated energies of excited states as well as the transition probabilities  are compared to the available experimental data showing a good overall agreement.  In addition, we present the RVAMPIR approach which provides a good description of the ground and gamma bands in the absence of strong mixing. 
\end{abstract}
\maketitle
\section{Introduction}\label{intro}
Self-consistent mean field methods with effective phenomenological interactions and their extensions beyond mean field provide the appropriate theoretical tools for describing many phenomena along the whole chart of nuclides, from light to medium, heavy and superheavy nuclei in or far away from the stability valley \cite{Bender_RMP_03}. On the one hand, the success of these methods is related to the high quality of the phenomenological effective interactions used -Skyrme, Gogny or Relativistic Mean Field (RMF). On the other hand, the mean field method allows the inclusion of many correlations within a very simple intrinsic product wave function. Hence, bulk properties such as masses and radii are very well described at the mean field level. However, in some cases this picture fails to take into account important correlations and methods beyond the mean field approach have to be applied. Furthermore, because the mean field is defined in the intrinsic frame it is mandatory to go beyond this approximation  to evaluate excitation energies or transition probabilities in the laboratory system.\\
There are several methods to incorporate the correlations missing at the mean field level. Normally, the intrinsic wave functions are allowed to break relevant symmetries of the system, for example,  particle number, rotational and translational invariance, parity, time-reversal, etc.  to enlarge the variational space and incorporate, for instance, deformation or superfluidity in the mean field picture. This leads to a degeneracy of the wave functions {\bf rotated} 
in the gauge space associated to the broken symmetry. An appropriated superposition of these wave functions
provides a symmetry conserving many-body wave function and an additional lowering of the energy of the system.
In this way, using projection techniques \cite{RingSchuck}, many correlations are obtained by restoring some or all of these symmetries . Furthermore, the mixing of different mean field configurations within the general framework of the Generator Coordinate Method (GCM) \cite{RingSchuck} allows the inclusion of quantum fluctuations along some relevant collective variables such as  the multipole moments.\\
Most of the currently used beyond mean field calculations with effective forces include two symmetry restoration, i.e., particle number (PN) and angular momentum projection (AMP) and configuration mixing along the {\bf axial} quadrupole deformation \cite{Bender_RMP_03,Rod_CaTiCr_07,Niksic_PNAMP_06}. This approach (axial GCM-PNAMP) has been successfully applied to study many phenomena like, for example, the appearance or degradation of shell closures in neutron rich nuclei \cite{Ray_Mg_04,Rod_Mg_09,Rod_CaTiCr_07,Jung_PLB_08}, shape coexistence in proton rich Kr \cite{Bender_Kr_06} or Pb \cite{Dug_Pb_03,Ray_Pb_04} isotopes or shape transitions in the $A\sim150$ region \cite{Niksic_PRL_07,Rod_PLB_08}. However, the intrinsic wave functions used there were restricted to have axial symmetry, with $K=0$, because this assumption simplifies considerably the angular momentum projection and lightens the computational burden significantly. This restriction is one of the major drawbacks of the method because it limits its applicability to systems where triaxiality does not play an important role. However, many exciting experimental and theoretical phenomena are closely related to the triaxial degree of freedom, for instance: the presence of $\gamma$-bands in the low lying energy spectra and $\gamma$-softness, shape coexistence and shape transitions in transitional regions \cite{Casten_O6_85, Casten_BaE5_00, Regan_ShapeTrans_03, Ita_10Be_02, Clem_Kr_07, Ober_Kr_09, Mo_96}; the lowering of fission barriers along the triaxial path \cite{Baran_Fiss_81, Bend_Fiss_98, Warda_Fiss_02}; the influence of triaxial deformation in the ground state for the mass models \cite{Naza_MassTriax_05, Moller_MassTriax_06}; triaxiality at high spin \cite{Naza_WS_85, Carl_142Gd_08, Yad_168Hf_08}; 
the observation of $K$-bands and isomeric states in the  Os region \cite{Os_85, Vik_Os_09, Kum_Os_09}; or some other exotic excitation modes such as wobbling motion and chiral bands \cite{Wobb_01, Chiral_04, Chiral_06}. \\
From the theoretical point of view some approaches beyond mean field have been proposed to study the triaxial effects. In particular, one of the most widely used is the collective Hamiltonian \cite{RingSchuck} given in different versions depending on the underlying nucleon-nucleon interaction used to define the collective potential, namely Pairing-plus-Quadrupole \cite{BKColl_PPQ_68}, Interacting Boson Model \cite{Cast_IBM_84}, Nilsson Woods-Saxon \cite{Naza_WS_85}, Gogny \cite{GGColl_Gogny_83,GDColl_Gogny_Kr_09, GOColl_Gogny_N40_09} or RMF \cite{Niksic_Rel_09}. This model has been applied successfully to describe some of the experimental features mentioned above. However, the collective Hamiltonian can be understood as a gaussian overlap approach (GOA) of the triaxial GCM and this description should be improved including properly the effects of the symmetry restoration and the full configuration mixing without any GOA approximation.\\
In the past, exact angular momentum projection with triaxial intrinsic wave functions without GCM have been  carried out only for schematic forces and/or reduced configuration spaces. Examples are  the projection of BCS \cite{RingBCS_PPQ_84} or Cranked Hartree-Fock-Bogoliubov (CHFB) states \cite{EnamiCHF_PPQ_99} with the Pairing-plus-Quadrupole interaction, the projection of Cranked Hartree-Fock (CHF) states (no pairing) with schematic \cite{HeenCHF_SK_84} and full Skyrme interactions \cite{DobaCHF_SK_07} or angular momentum projection before variation with particle number and parity restoration in limited shell model spaces  \cite{VAMPIR,MONSTER} .\\
However, recent improvement of the computational capabilities enabled the first implementations  of the angular momentum projection of triaxial intrinsic wave functions in the whole $(\beta,\gamma)$ plane with effective forces. In particular, Bender and Heenen reported GCM calculations with particle number and triaxial angular momentum projection (PNAMP) with the Skyrme SLy4 interaction \cite{BendHFB_SK_08}. In this work, the intrinsic wave functions were found by solving the Lipkin-Nogami (LN) equations. On the other hand, Yao \textit{et al} presented the implementation of the triaxial angular momentum projection \cite{RingAMP_Rel_09} and the extension to the GCM  \cite{RingGCM_Rel_10} for the Relativistic Mean Field (RMF) framework. In the latter work  no particle number projection has been performed and the mean field states are found by solving the RMF+BCS instead of the full HFB or LN equations. These approximations could lead to a poor description of important pairing correlations, especially in the weak pairing regime where even spurious phase transitions appear \cite{Anguiano_VAP_02,Rod_CaTiCr_07}.\\
In this paper we present the first implementation  of the Generator Coordinate Method with Particle Number and Angular Momentum Projected (GCM-PNAMP) triaxial HFB wave functions with the finite range density dependent Gogny force \cite{BERGNPA84}.   The finite range of the Gogny force provides excellent  pairing properties and is often used as a benchmark in this respect. Furthermore it is able to provide  at the same time both good global  as well as spectroscopic properties \cite{egi_Pb_04,egi_NoFission_00}.
The intrinsic HFB states are found by solving the Variation After Particle Number Projection (VAP-PN) equations \cite{Anguiano_VAP_01}. This fact constitutes the main methodological difference with respect to the calculations reported in Ref. \cite{BendHFB_SK_08}. This is a very important difference because VAP-PN allows the inclusion of the pairing correlations in a very efficient way yielding a significant improvement of the final results with respect to other approaches \cite{Anguiano_VAP_01,Rod_CaTiCr_07}. \\
    In nuclei without  strong mixing  the  so called Variation After Mean-field Projection In Realistic  model spaces (VAMPIR) \cite{VAMPIR}  approach  has been very successful. In this approach only one HFB wave function is considered which is
  determined by minimization of the projected energy, i.e. the VAP approach for both the AM and the PN projections.  A full VAMPIR approach with the Gogny force and large configuration spaces is not feasible yet. Instead we use an approximation to it,  which we call RVAMPIR  and which as we shall see, in the case nucleus studied in this article, provides very reasonable
  results  for the ground and $\gamma$ bands  with much less effort than in the GCM approach. \\
The paper is organized as follows. In Sec. \ref{theo_frame1} we will give an overview of the theoretical framework. Then, we will focus our analysis on the nucleus  $^{24}$Mg which has been studied as a test case in earlier implementations of the GCM-PNAMP method with Skyrme and Relativistic interactions. In particular, in Sec. \ref{axial_calc} we will show a standard axially symmetric  calculation 
which allows to make an educated guess for some relevant parameters needed in the full calculation such as the number of major oscillator shells or the relevant deformations ranges. In Sec. \ref{PNAMP} we will analyze in detail the simpler PNAMP method, studying the convergence of the integrals in the Euler angles, giving some consistency requirements and showing the role of having an adequate mesh in the $(\beta,\gamma)$ plane. In Sec. \ref{GCM-PNAMP} we will show the final results for the calculated spectrum and B(E2) transitions strengths of $^{24}$Mg and a comparison with experimental data. Finally, a brief summary and outlook on future work will be addressed in Sec. \ref{summary}.  
\section{Theoretical framework}\label{theo_frame1}
\subsection{Generator Coordinate Method with Particle Number and Angular Momentum Projected states (GCM-PNAMP)}\label{theo_frame}
In the present approach, the final many-body wave functions that describe the different states of an even-even nucleus with $Z(N)$ number of protons (neutrons) are written as:
\begin{equation}
|IM;NZ \sigma\rangle=\sum_{K\beta\gamma}f^{I;NZ,\sigma}_{K\beta\gamma}|IMK;NZ;\beta\gamma\rangle
\label{gcm_state}
\end{equation}   
where ($\beta,\gamma$) are quadrupole deformation parameters (see below), $\sigma=1,2,...$ labels the levels for a given value of the angular momentum $I$ and $M,K$ are the projections of $\vec{I}$ on the laboratory and intrinsic $z-$axes respectively. The coefficients  
$f^{I;NZ,\sigma}_{K\beta\gamma}$ of the linear combination are found by minimizing the energy within the non-orthogonal set of wave functions $\lbrace|IMK;NZ;\beta\gamma\rangle\rbrace$. These  states are obtained  by projecting the intrinsic mean-field states $|\Phi(\beta,\gamma)\rangle$ onto good particle number and angular momentum:
\begin{equation}
|IMK;NZ;\beta\gamma\rangle=\frac{2I+1}{8\pi^{2}}\int\mathcal{D}^{I*}_{MK}(\Omega)\hat{R}(\Omega)\hat{P}^{N}\hat{P}^{Z}|\Phi(\beta,\gamma)\rangle d\Omega
\label{IMK_WF}
\end{equation}
with $\hat{P}^{N}=\frac{1}{2\pi}\int_{0}^{2\pi}e^{i\varphi(\hat{N}-N)}d\varphi$ the neutron number projector ( $\varphi$ the associated gauge angle and  $\hat{P}^{Z}$ protons the proton number projector),  $\hat{R}(\Omega)$ and $\mathcal{D}^{I*}_{MK}(\Omega)$ are the rotation operator and the Wigner matrices    
 \cite{Varsha} in the Euler angles $\Omega=(a,b,c)$  \footnote{We choose for the Euler angles the notation $\Omega=(a,b,c)$ instead of the usual $(\alpha,\beta,\gamma)$ to avoid confusion with the deformation parameters $(\beta,\gamma)$}, respectively. In principle, the ranges for these angles are $(0\leq a\leq 2\pi,0\leq b\leq \pi, 0\leq c\leq 2\pi)$. However, for intrinsic  Hartree-Fock-Bogoliubov (HFB)  states ($|\Phi(\beta,\gamma)\rangle$) which are symmetric under time-reversal and simplex symmetries, the intervals for both gauge and Euler angles can be reduced to $(0\leq \varphi\leq \pi/2)$ and $(0\leq a\leq \pi/2,0\leq b\leq \pi/2, 0\leq c\leq \pi)$, respectively \cite{BendHFB_SK_08}.\\
The wave functions (Eq. \ref{IMK_WF}) are eigenstates of the particle number and angular momentum operators:
\begin{eqnarray}
\hat{N}|IMK;NZ;\beta\gamma\rangle&=& N|IMK;NZ;\beta\gamma\rangle, \\
\hat{Z}|IMK;NZ;\beta\gamma\rangle&=&Z|IMK;NZ;\beta\gamma\rangle,\\
\hat{I}^{2}|IMK;NZ;\beta\gamma\rangle&=&\hbar^{2}I(I+1)|IMK;NZ;\beta\gamma\rangle\label{I2}\\
\hat{I}_{z}|IMK;NZ;\beta\gamma\rangle&=&\hbar M|IMK;NZ;\beta\gamma\rangle,\\
\hat{I}_{3}|IMK;NZ;\beta\gamma\rangle&=&\hbar K|IMK;NZ;\beta\gamma\rangle
\end{eqnarray}
\\ 
The intrinsic HFB  states ($|\Phi(\beta,\gamma)\rangle$) are obtained by minimizing the particle-number  projected energy functional $E^{N,Z}\left[\bar{\Phi}(\beta,\gamma)\right]$  (variation after projection, VAP)\cite{Anguiano_VAP_01}. This is one of the most relevant parts in the calculation because the quality of the result largely depends on the structure of the intrinsic HFB-type wave functions used. In contrast to other methods like plain HFB or Projected Lipkin-Nogami (PLN), the VAP-PN performs the restoration of the particle number symmetry in an optimal way, including pairing correlations both in the weak and strong pairing regimes \cite{Rod_CaTiCr_07}.
 This is especially relevant in GCM-like theories where a large grid of $(\beta,\gamma)$ points is needed.  The strength of the pairing correlations has a strong dependence  on the single particle level density and the latter one itself with the deformation parameters. This implies that a strongly $(\beta,\gamma)$ dependent oscillating pairing regime appears in the calculations and consequently theories like plain HFB (BCS) or PLN (LN) are unable to cope with this challenge providing wave functions of oscillating goodness. Only a VAP-PN approach warrants  high quality solutions  independently  of the $(\beta,\gamma)$ values.\\
 Dealing with effective forces like Skyrme, Relativistic and Gogny,  a natural separation
of the interaction into  the two-body Hamiltonian $\hat{H}_{\rm 2b}$  on the one hand and the density-dependent part,
$\varepsilon^{N,Z}_{DD}[\Phi]$ on the other emerges.
In our case, we are using the Gogny D1S interaction \cite{BERGNPA84} and $\hat{H}_{\rm 2b}$ corresponds to the kinetic energy
(the two-body part from the center of mass correction included)  plus the spin-orbit, Coulomb and the finite range central potentials. In the calculations, all direct, exchange and pairing terms are included \cite{Anguiano_EXC_01}.  The VAP-PN principle, provides
\begin{equation}
\delta  \biggl. \displaystyle E^{N,Z}\left[\bar{\Phi}(\beta,\gamma)\right] \biggr|_{\bar{\Phi}=\Phi} =0
\label{vap_min}
\end{equation}
where:
\begin{equation}
E^{N,Z}[\Phi]=\frac{\langle\Phi|\hat{H}_{\rm 2b}\hat{P}^{N}\hat{P}^{Z}|\Phi\rangle}{\langle\Phi|\hat{P}^{N}\hat{P}^{Z}|\Phi\rangle}+\varepsilon^{N,Z}_{DD}[\Phi]-\lambda_{q_{20}}\langle\Phi|\hat{Q}_{20}|\Phi\rangle-\lambda_{q_{22}}\langle\Phi|\hat{Q}_{22}|\Phi\rangle
\label{vap_funct}
\end{equation}
 In a beyond mean field method, and in particular for the  particle number projection, we need a reasonable prescription for the spatial density, which we shall call $\rho_{\rm int}(\vec{r})$, that enters in $\varepsilon^{N,Z}_{DD}[\Phi]$, the density dependent term of the interaction.
In this work, assuming the phenomenological nature of these interactions and considering that the restoration of the particle number symmetry is performed not in the coordinate  but in the gauge space, we have chosen the number projected spatial density prescription that has proven to be free of  divergences \cite{Anguiano_VAP_01} and to give very good results for describing many phenomena along the nuclear chart:
\begin{equation}
\rho^{NZ}_{\rm int}(\vec{r})\equiv\frac{\langle\Phi|\hat{\rho}(\vec{r})P^{N}P^{Z}|\Phi\rangle}{\langle\Phi|P^{N}P^{Z}|\Phi\rangle}
\end{equation}
with $\hat{\rho}(\vec{r})\equiv\int d\vec{r}' \delta(\vec{r}-\vec{r}')$.  As shown  in \cite{Anguiano_VAP_01} for the PNP  and in \cite{Valor97} for the Lipkin-Nogami approach,  the use of the projected density or the so-called mixed prescription  (in the case when the latter  is free of potential divergences) provide very similar results. 
Furthermore, we see in Eq.  \ref{vap_funct} that the minimization is performed under constraints on the quadrupole deformation operators $\hat{Q}_{2\mu}$. The Lagrange multipliers $\lambda_{q_{2\mu}}$ ensure that the following conditions are fulfilled in the intrinsic state:
\begin{eqnarray}
\lambda_{q_{20}}\rightarrow\langle\Phi|\hat{Q}_{20}|\Phi\rangle=q_{20}\nonumber\\
\lambda_{q_{22}}\rightarrow\langle\Phi|\hat{Q}_{22}|\Phi\rangle=q_{22}
\end{eqnarray}
In addition, the deformation parameters $(\beta,\gamma)$ are directly related to $(q_{20},q_{22})$ by:
\begin{eqnarray}
q_{20}=\frac{\beta\cos\gamma}{C}\,\,;\,\,q_{22}=\frac{\beta\sin\gamma}{\sqrt{2}C}\,\,;\,\,C=\sqrt{\frac{5}{4\pi}}\frac{4\pi}{3r_{0}^{2}A^{5/3}}
\label{betagamma}
\end{eqnarray}
being $r_{0}=1.2$ fm and $A$ the mass number. These constraints allow to explore the  $(\beta,\gamma)$ plane to
generate the wave functions to be used in the  configuration mixing calculations.\\
We now describe the Generator Coordinate Method (GCM) to obtain the final spectrum ($E^{I;NZ;\sigma}$) and the coefficients $f^{I;NZ,\sigma}_{K\beta\gamma}$ given in Eq. \ref{gcm_state}.  Minimization of the energy with respect
to the coefficients $f^{I;NZ,\sigma}_{K\beta\gamma}$ leads to  the Hill-Wheeler-Griffin (HWG) equation 
\begin{equation}
\sum_{K'\beta'\gamma'}\left(\mathcal{H}^{I;NZ}_{K\beta\gamma K'\beta'\gamma'}-E^{I;NZ;\sigma}\mathcal{N}^{I;NZ}_{K\beta\gamma K'\beta'\gamma'}\right)f^{I;NZ;\sigma}_{K'\beta'\gamma'}=0,
\label{HW_eq}
\end{equation}
which has to be solved for each value of the angular momentum.  The GCM norm- and energy-overlaps  have been defined as:
\begin{eqnarray}
\mathcal{N}^{I;NZ}_{K\beta\gamma K'\beta'\gamma'}&\equiv&\langle IMK;NZ;\beta\gamma|IMK';NZ;\beta'\gamma'\rangle\nonumber\\
\mathcal{H}^{I;NZ}_{K\beta\gamma K'\beta'\gamma'}&\equiv&\langle IMK;NZ;\beta\gamma|\hat{H}_{\rm 2b}|IMK';NZ;\beta'\gamma'\rangle+\varepsilon^{IKK';NZ}_{DD}\left[\Phi(\beta,\gamma),\Phi'(\beta',\gamma')\right]
\label{gcm_overlaps}
\end{eqnarray}
In the last expression, we have separated again the energy overlap in the contribution of the pure Hamiltonian part of the interaction and the density-dependent term. In the latter, we have used the particle number projected spatial density combined with the mixed prescription for the angular momentum projection and GCM part, namely:
\begin{equation}
\rho^{NZ}_{\rm int}(\Omega,\vec{r})\equiv\frac{\langle\Phi|\hat{\rho}(\vec{r})\hat{R}(\Omega)P^{N}P^{Z}|\Phi'\rangle}{\langle\Phi|\hat{R}(\Omega)P^{N}P^{Z}|\Phi'\rangle}.
\end{equation}
This prescription is suitable for dealing with the restoration of broken symmetries  in the coordinate space such as the rotational invariance or the spatial parity.\\
Once we have calculated the corresponding GCM overlaps, the next step consists in solving the HWG equations (Eq. \ref{HW_eq}). 
To cope with the problem of the linear dependence one first introduces a orthonormal basis defined by the eigenvalues $n^{I;NZ}_{\Lambda}$ and eigenvectors $u^{I;NZ}_{K\beta\gamma;\Lambda}$ of the norm overlap:
\begin{equation}
\sum_{K'\beta'\gamma'}\mathcal{N}^{I;NZ}_{K\beta\gamma K'\beta'\gamma'} u^{I;NZ}_{K'\beta'\gamma';\Lambda}=n^{I;NZ}_{\Lambda}u^{I;NZ}_{K\beta\gamma;\Lambda}.
\label{zeta}
\end{equation}
This orthonormal basis is known as the natural basis and  for $n^{I;NZ}_{\Lambda}$ values such that $n^{I;NZ}_{\Lambda}/n^{I,NZ}_{max}>\zeta$, the  natural states are defined by:
\begin{equation}
|\Lambda^{IM;NZ}\rangle=\sum_{K\beta\gamma}\frac{u^{I;NZ}_{K\beta\gamma;\Lambda}}{\sqrt{n^{I;NZ}_{\Lambda}}}|IMK;NZ;\beta\gamma\rangle.
\label{natstates}
\end{equation}
Obviously, a cutoff  $\zeta$ in the value  of the norm eigenvalues has to be introduced in order to avoid linear dependences \cite{RingAMP_Rel_09}.
Then, the HWG equation is transformed into a normal eigenvalue problem:
\begin{equation}
\sum_{\Lambda'}\langle\Lambda^{I;NZ}|\hat{H}|\Lambda'^{I;NZ}\rangle G^{I;NZ;\sigma}_{\Lambda'}=E^{I;NZ;\sigma}G^{I;NZ;\sigma}_{\Lambda}.
\end{equation}
From the coefficients $G^{I;NZ;\sigma}_{\Lambda}$ we can define the so-called collective wave functions $F^{I;NZ;\sigma}(\beta,\gamma)$ that account for the  probability density, normalized to 1,  of finding the state $(I,\sigma)$ with given deformation parameters $(\beta,\gamma)$:
\begin{equation}
F^{I;NZ;\sigma}(\beta,\gamma)=\sum_{\Lambda,K}G^{I;NZ;\sigma}_{\Lambda}u^{I;NZ}_{K\beta\gamma;\Lambda}=
\sum_{K} F^{I;NZ;\sigma}_{K}(\beta,\gamma).
\label{coll_wf}
\end{equation} 
we have also introduced   $F^{I;NZ;\sigma}_{K}(\beta,\gamma)$ that account for the  probability density of finding the state $(I,\sigma)$ with given values of $K$ and deformation parameters $(\beta,\gamma)$.

 Furthermore, the expectation value of a generic operator $\hat{O}$ is given by
\begin{equation}
o^{I;NZ;\sigma}=\sum_{\Lambda;\Lambda'}\sum_{K\beta\gamma;K'\beta'\gamma'}
G^{I;NZ;\sigma*}_{\Lambda}\frac{u^{I;NZ*}_{K\beta\gamma;\Lambda}}{\sqrt{n^{I;NZ}_{\Lambda}}}\langle \varpi|\hat{O}|\varpi'\rangle\frac{u^{I;NZ}_{K'\beta'\gamma';\Lambda'}}{\sqrt{n^{I;NZ}_{\Lambda'}}}G^{I;NZ;\sigma}_{\Lambda'},
\end{equation}
with $\langle \varpi|\hat{O}|\varpi'\rangle =\langle IMK;NZ;\beta\gamma|\hat{O}|IMK';NZ;\beta'\gamma'\rangle$. This expression  can be  generalized to account for transitions associated to the tensorial operator $\hat{T}_{12}$:
\begin{equation}
t(I_{1}\sigma_{1}\rightarrow I_{2}\sigma_{2})=\sum_{\Lambda;\Lambda'}\sum_{K\beta\gamma;K'\beta'\gamma'}
G^{I_{1};NZ;\sigma^{*}_{1}}_{\Lambda}\frac{u^{I_{1};NZ*}_{K\beta\gamma;\Lambda}}{\sqrt{n^{I_{1};NZ}_{\Lambda}}}\langle \varpi_1||\hat{T}_{12}||\varpi_{2}'\rangle\frac{u^{I_{2};NZ}_{K'\beta'\gamma';\Lambda'}}{\sqrt{n^{I_{2};NZ}_{\Lambda'}}}G^{I_{2};NZ;\sigma_{2}}_{\Lambda'},
\label{trans_pro}
\end{equation}
where $\langle \varpi_1 ||\hat{T}_{12}|| \varpi_{2}'\rangle=\langle I_{1}K;NZ;\beta\gamma||\hat{T}_{12}||I_{2}K';NZ;\beta'\gamma'\rangle$ stands for the reduced matrix element calculated according to the Wigner-Eckart theorem \cite{RingSchuck,Varsha}. Detailed expressions for calculating these reduced matrix elements for B(E2) transitions and spectroscopic quadrupole moments within this framework can be found elsewhere \cite{BendHFB_SK_08,RingGCM_Rel_10,Ray_Mg_04}.
\subsection{Simpler approaches: Particle Number and Angular Momentum Projection (PNAMP) and the RVAMPIR approximation} \label{RVAMPIR}
The expressions given above constitute the most general framework that we are using for solving the nuclear many body problem. Nevertheless, there are some limiting cases with a relevant physical meaning that can be deduced  in a straightforward manner from them. The first one is the particle number projection (PNP) that has been discussed above (Eq. \ref{vap_funct}). The second approach is the particle number and angular momentum projection (PNAMP) of a single point in the $(\beta,\gamma)$ plane. Here, the wave function is of the form of Eq. \ref{gcm_state} but without the mixing in the deformation parameters:
\begin{equation}
|IM;NZ;\nu;\beta,\gamma\rangle=\sum_{K}h^{I;NZ,\nu}_{K}(\beta\gamma)|IMK;NZ;\beta\gamma\rangle,
\label{pnamp_state}
\end{equation}
where the label $\nu$ stands for the $(2I+1)$ different states that can be obtained with the angular momentum projection. However, due to the time reversal and simplex symmetries imposed on  the intrinsic wave functions, this number is reduced to $(I/2+1)$ and $((I-1)/2)$ states for even and odd values of $I$, respectively. Moreover, if we furthermore have axial symmetry, only one state can be obtained and
only  for even values of $I$.\\
The coefficients $h^{I;NZ,\nu}_{K}(\beta\gamma)$ and the PNAMP energies $E^{I;NZ;\nu}(\beta,\gamma)$ are found by solving the simplified version of the HWG equation (see Eq. \ref{HW_eq}):
\begin{equation}
\sum_{K}\left(\mathcal{H}^{I;NZ}_{K\beta\gamma K'\beta\gamma}-E^{I;NZ;\nu}(\beta,\gamma)\mathcal{N}^{I;NZ}_{K\beta\gamma K'\beta\gamma}\right)h^{I;NZ;\nu}_{K'}(\beta,\gamma)=0.
\label{HW_eq_pnamp}
\end{equation}
The remaining expressions used to solve the HWG equations are simplified in the same manner, i.e., removing the sum over $(\beta,\gamma)$ from the equations and evaluating only the diagonal part. In addition, the collective wave functions $F^{I;NZ;\nu}_{K}(\beta,\gamma)$ (Eq. \ref{coll_wf}), which in analogy we shall call  $H^{I;NZ;\nu}_{K}(\beta,\gamma)$,  now give  the spectral distribution in the $K$ space of the corresponding PNAMP state. \\
  A full variation of the HFB wave function in the VAP approach, in the spirit of  VAMPIR, for the PN {\bf and} the AM with large configuration spaces and the Gogny interaction is not yet feasible.  
However we can use an approximation to  VAMPIR, which we shall call from now on RVAMPIR, in which  the PN  is handled in the VAP approach and the AM in a Restricted VAP (RVAP) one.  The RVAP approximation has been thoroughly studied  in  \cite{TOM}-\cite{SCHUNCK}. In the VAP method the whole Hilbert space associated with the HFB transformation is scanned  in the variational procedure. In the RVAP method, however, only a restricted variational space of highly correlated wave-functions is allowed in the minimization process.  Monopole (pairing) and quadrupole ($\beta$ and $\gamma $)  correlations are believed to be the most relevant degrees of freedom of atomic nuclei and are related to the particle number and the angular momentum symmetries, respectively.    
  Since we are considering the PN symmetry in the VAP theory it seems reasonable in our case to consider  the restricted Hilbert space to contain  a whole set of quadrupole deformed wave-functions $|\Phi(\beta,\gamma)\rangle$ which parametrically depend on  $(\beta,\gamma)$.  This procedure is justified by  theoretical arguments \cite{RingSchuck}  which establish that a VAP approach is needed for systems with weakly broken symmetries, like in the PN case where only a few Cooper pairs participate,  but it can be approximated in case of strongly broken symmetries, such as deformation, where a large number of nucleons participate. Concerning the differences  of this approximation as compared to  VAMPIR it is clear \cite{TOM} that if, besides of considering the quadrupole moments
 $\hat{Q}_{20}$ and  $\hat{Q}_{22}$  in Eq.~\ref{vap_funct}, we will include higher multipole moments $\hat{Q}_{LM}$  to increase
 the variational space, our solution would get very close to the one of the genuine VAMPIR.  With respect of the quality of our approach (again with respect to the full VAMPIR) we expect that in general it will be very similar and only in very soft nuclei, where higher modes (hexadecupole for example) are very relevant, differences may arise. But for very soft nuclei we have to question also the full VAMPIR since a GCM-like approach will be more appropriate. That means, RVAMPIR is not as ``restricted"  as its name might imply.\\
Specifically the basic RVAMPIR approach  consist of the following steps:
\begin{itemize}
\item[A.-] At {\bf each} $(\beta_i,\gamma_i)$ value  of a given set  of points  in the $(\beta,\gamma)$ plane the following items are performed:
\begin{itemize}
\item[A1.-] Solve the VAP-PN equations, Eqs.\ref{vap_min}-\ref{vap_funct}, to determine the $\beta-\gamma$ constrained HFB wave function $|\Phi(\beta,\gamma)\rangle$.
\item[A2.-] Carry out simultaneous  particle number and angular momentum  projection  on the wave function  $|\Phi(\beta,\gamma)\rangle$, what we have called $|IMK; NZ;\beta\gamma \rangle$, see Eq.~\ref{IMK_WF}, to form  the  linear combination of the state  $|IM; NZ;\nu; \beta\gamma \rangle$  in Eq.~\ref{pnamp_state}.
\item[A3.-] Solve the HWG equation, Eq.\ref{HW_eq_pnamp},  for different angular momenta.
\end{itemize}
\item[B.-] For {\bf each} value of the angular momentum sort out the energies $E^{I;NZ;\nu}(\beta,\gamma)$ of
 Eq.~\ref{HW_eq_pnamp} and find out the point $(\beta^{I}_{\rm min},\gamma^{I}_{\rm min})$ providing the energy minimum $E^{I;NZ;\nu}_{\rm min}(\beta^{I}_{\rm min},\gamma^{I}_{\rm min})$
\item[C.-] The solutions of the HWG equation at the points $(\beta^{I}_{\rm min},\gamma^{I}_{\rm min})$ provide $I/2+1 ((I-1)/2)$  states for even (odd) $I$ values, which allow to build  a partial spectrum  and  to calculate the transition probabilities among the different states or any other observable. 
 
   One has to notice that all RVAMPIR states are orthogonal, those with different AM in an obvious way and those
 with the same AM because they are solution of the same eigenvalue equation.
\end{itemize}

 In the following sections we will give some examples of the convergence, consistency and performance of the methods described above. All the many body intrinsic wave functions and operators have been expanded in a cartesian harmonic oscillator single particle basis closed under rotations \cite{egido_rot_93}. In particular, the rotation operator $\hat{R}(\Omega)$ has been evaluated following the expressions given in Ref. \cite{Nad_RotMat} and the Neergard method \cite{Neer_Normover} has been used in the calculation of the norm overlaps in order to determine the correct sign of the Onishi formula \cite{Onishi,BalianBrezin,Hara}.  The overlaps of a generic operator have been calculated using the generalized Wick theorem \cite{BalianBrezin}.  \\
\section{Axial calculations for $^{24}$Mg}
\label{axial_calc}
Due to the huge computational cost of the full triaxial calculation, it is important to study first the axial case (with $K=0$) in order to fix some relevant quantities. The most important ones are the region of $\beta$ deformation to be included in the calculation and the number of major oscillator shells in which the mean field wave functions are expanded. The computational effort depends critically on these quantities and it is important to ensure the convergence of the results, at least in the axial case,  to have  a reasonable choice 
which then later allows to perform the full triaxial calculation.\\
The main advantage of considering only axial symmetric ($K=0$) intrinsic wave functions $|\Phi(\beta,\gamma=0^{\circ},180^{\circ})\rangle\equiv|\Phi(\beta)\rangle$ is that the integration over the Euler angles $(a,c)$ can be done analytically and this fact reduces drastically the computational time. The simplified expressions of the axial GCM-PNAMP method can be found in detail in Ref. \cite{Ray_Mg_04}. 
\begin{figure}[t]
\centering
\includegraphics[width=0.4\textwidth]{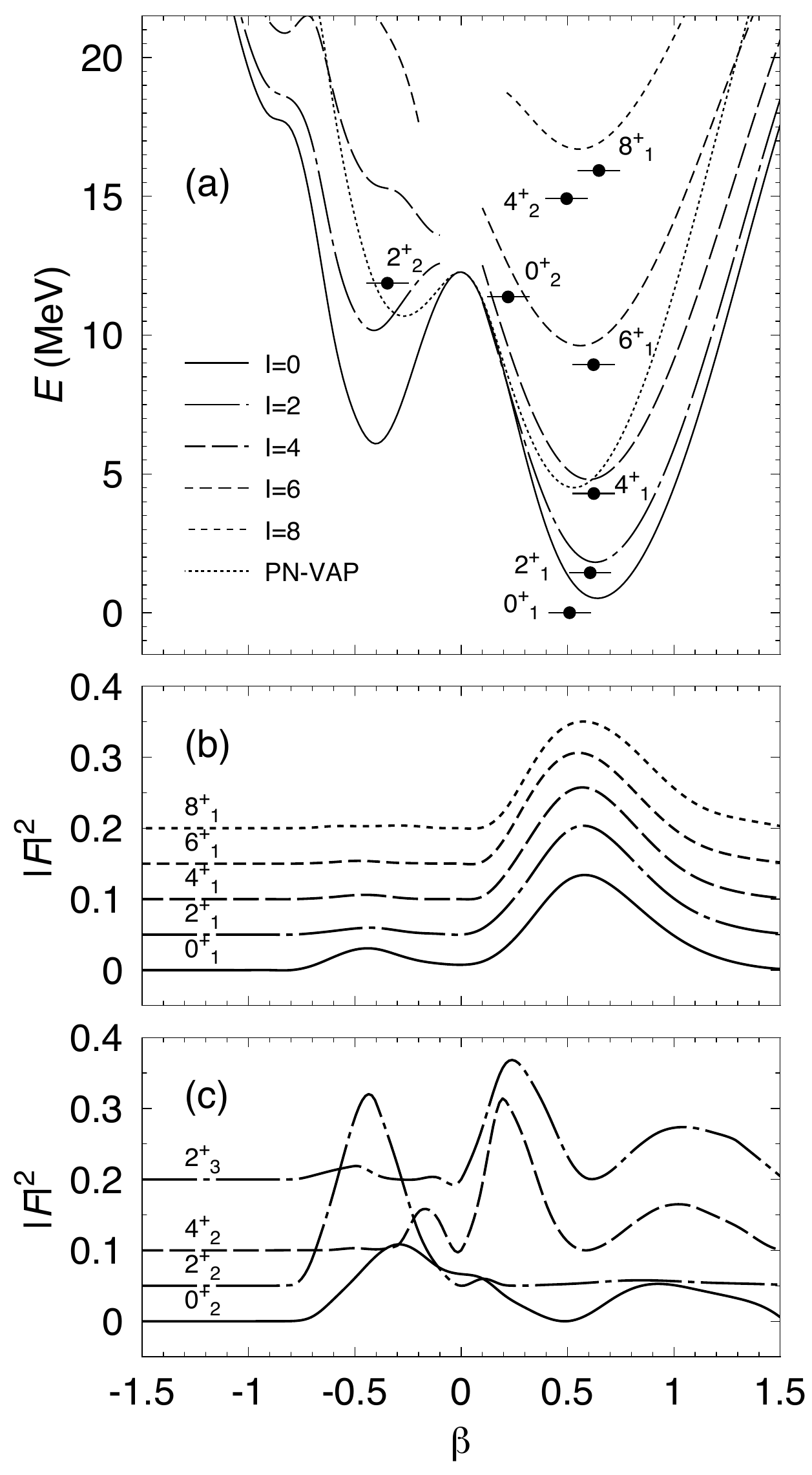}
\caption{(a) Potential Energy Surfaces (PES) along the $\beta$ deformation for particle number projection and particle number and angular momentum projection ($^{24}$Mg). The bullets correspond to the excitation energies for the different GCM levels $(I,\sigma)$ with their positions at $\bar{\beta}^{I\sigma}=\sum_{\beta}\beta|F^{I\sigma}(\beta)|^{2}$. The energy is normalized to the GCM ground state energy ($0^{+}_{1}$). (b) Collective wave functions for the $(\sigma=1)$ GCM levels. The values of the ordinate axis is displaced by 0.05 with increasing angular momentum. (c) Same as (b) but for the $(\sigma=2)$ GCM levels and $2^{+}_{3}$ state. Positive and negative values of $\beta$ correspond to prolate ($\gamma=0^{\circ}$) and oblate ($\gamma=180^{\circ}$) shapes, respectively.}
\label{Fig1}
\end{figure}
We first analyze  the results obtained for the nucleus  $^{24}$Mg using $N_{shells}=7$ oscillator shells and $N_{points}=31$ intrinsic wave functions distributed in the interval $(-1.5\leq\beta \leq1.5)$ with positive and negative values of $\beta$ corresponding to prolate $\gamma=0^{\circ}$ and oblate $\gamma=180^{\circ}$ shapes. The integration over the gauge angle $\varphi$ for the particle number projection part has been performed using the Fomenko expansion \cite{Fomenko} while for the integration over the Euler angle $b$ a Gaussian-Legendre quadrature has been used. We have chosen $N_{Fom}=9$ and $N_{b}=16$ as the number of integration points for the particle number and the angular momentum parts of the projection, respectively. With these assumptions the expectation values for the $\hat{N}$, $\hat{Z}$, $\hat{N}^{2}$, $\hat{Z}^{2}$ and $\hat{I}^{2}$ operators differ by  less than $10^{-8}$ from the corresponding eigenvalues. In Fig. \ref{Fig1}(a) we plot the potential energy surfaces (PES) along the $\beta$ direction for the VAP-PN and PNAMP approaches. The VAP-PN curve shows two differentiated minima separated by a barrier of $\sim 7.7$ MeV, the lowest one at prolate deformation ($\beta=0.5$) and the other one in the oblate part ($\beta=-0.2$) . These minima are shifted towards larger  deformations when the angular momentum projection is performed. In particular, a well defined prolate minimum appears at $\beta=0.6$ for $I=0,2,4,6,8$ showing that a rotational band will develop from this intrinsic state. For the ground state the gain in correlation energy  due to the restoration of the rotational symmetry amounts to  $\sim 3.9$ MeV. In the oblate part, we observe a minimum at $\beta=-0.4$ for $I=0,2$ while for higher values of the angular momentum the minimum vanishes. Here, the energy difference between the VAP-PN and the $I=0$ oblate minima is $\sim 4.6$ MeV.\\
The next step in the calculation is the configuration mixing of the PNAMP states. Hence, once the HWG equations are solved, we select as the final solutions those that belong to a \textit{plateau} in the energy as a function of the number of states in the natural basis (Eq. \ref{natstates}) and fulfill the orthonormality condition.  To avoid duplications a detailed
discussion on these issues is postponed to the triaxial case. The resulting GCM-PNAMP energies are also represented in Fig. \ref{Fig1}(a), while the corresponding collective wave functions (Eq. \ref{coll_wf}) are plotted in Fig. \ref{Fig1}(b) for $\sigma=1$ and in Fig. \ref{Fig1}(c)  for $\sigma=2$ and $2^+_3$.  In these figures we can see that the $\sigma=1$ states are members of a rotational band, with most of the intensity of the collective wave functions concentrated around $\beta=0.6$. This deformation corresponds to the location of the prolate minima of the different potential wells. The ground state $0^{+}_{1}$ also has a small mixing with the oblate minimum at $\beta=-0.5$. The situation is rather different for the $\sigma=2$ states. The second $0^{+}$ state is a mixing of oblate and prolate configurations, while wave function of the $2^{+}_{2}$ state peaks in the oblate minimum of the corresponding PES and the $4^{+}_{2}$ state could be considered as a vibration built on the  $I=4$ prolate well with a small contribution of slightly oblate states.
In the  $2^+_3$ state the prolate deformations are again favored. 
\begin{figure}[t]
\centering
\includegraphics[width=0.4\textwidth]{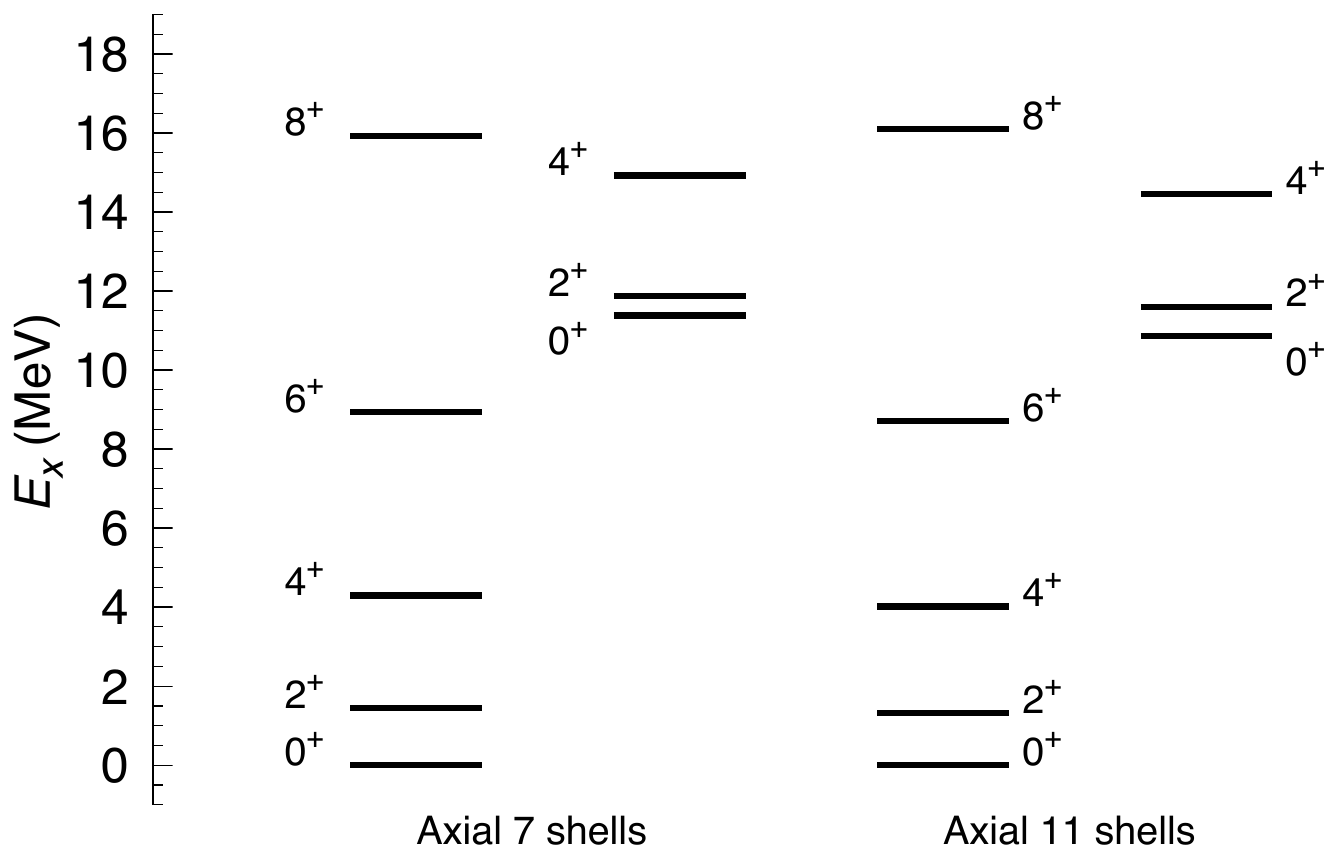}
\caption{Axial PNAMP-GCM excitation spectra  of $^{24}$Mg obtained considering 7 shells (left) and 11 shells (right) in the calculations.}
\label{Fig2}
\end{figure}
Remembering  that the purpose of this axial calculation is to determine the range of values of $\beta$ needed to obtain converged results in the low-lying energy spectrum, we observe in Fig. \ref{Fig1}(b)-(c) that all the collective wave functions studied here drop to zero at the boundaries. A smaller interval, however, could not be sufficient for describing correctly the collective states.  In addition, we have checked the convergence of the results as a function of the number of the points included in the GCM-PNAMP. Increasing this number  to $N_{points}=61$ still yields very similar results  for the PES, GCM-PNAMP energies and the collective wave functions as compared to the ones obtained for $N_{points}=31$.\\ Finally, in order to test the convergence with the number of oscillator shells, we have performed a calculation with $N_{shells}=11$ and $N_{points}=31$. It is noteworthy  that for a triaxial calculation, the computational time for $N_{shells}=11$ is $\sim$30 times larger than the one used for $N_{shells}=7$. Although this fact complicates the applicability of this method for heavy nuclei, for lighter systems the calculation with a smaller number of oscillator shells could 
still be sufficient. This is the case for $^{24}$Mg, where the PES and the collective wave functions calculated with $N_{shells}=11$  (not shown) are very similar to the $N_{shells}=7$ results.  In Fig. \ref{Fig2} we  compare the spectra obtained in the two calculations and observe a relative error of less than  10\% for all the levels.  While the members of the $\sigma=1$ bands almost match each other, small differences are found in the $\sigma=2$ band. This comparison justifies that  all further calculations are performed  with$N_{shells}=7$.
\section{Convergence and consistency of the triaxial PNAMP}\label{PNAMP}
\begin{figure}[t]
\centering
\includegraphics[width=0.6\textwidth]{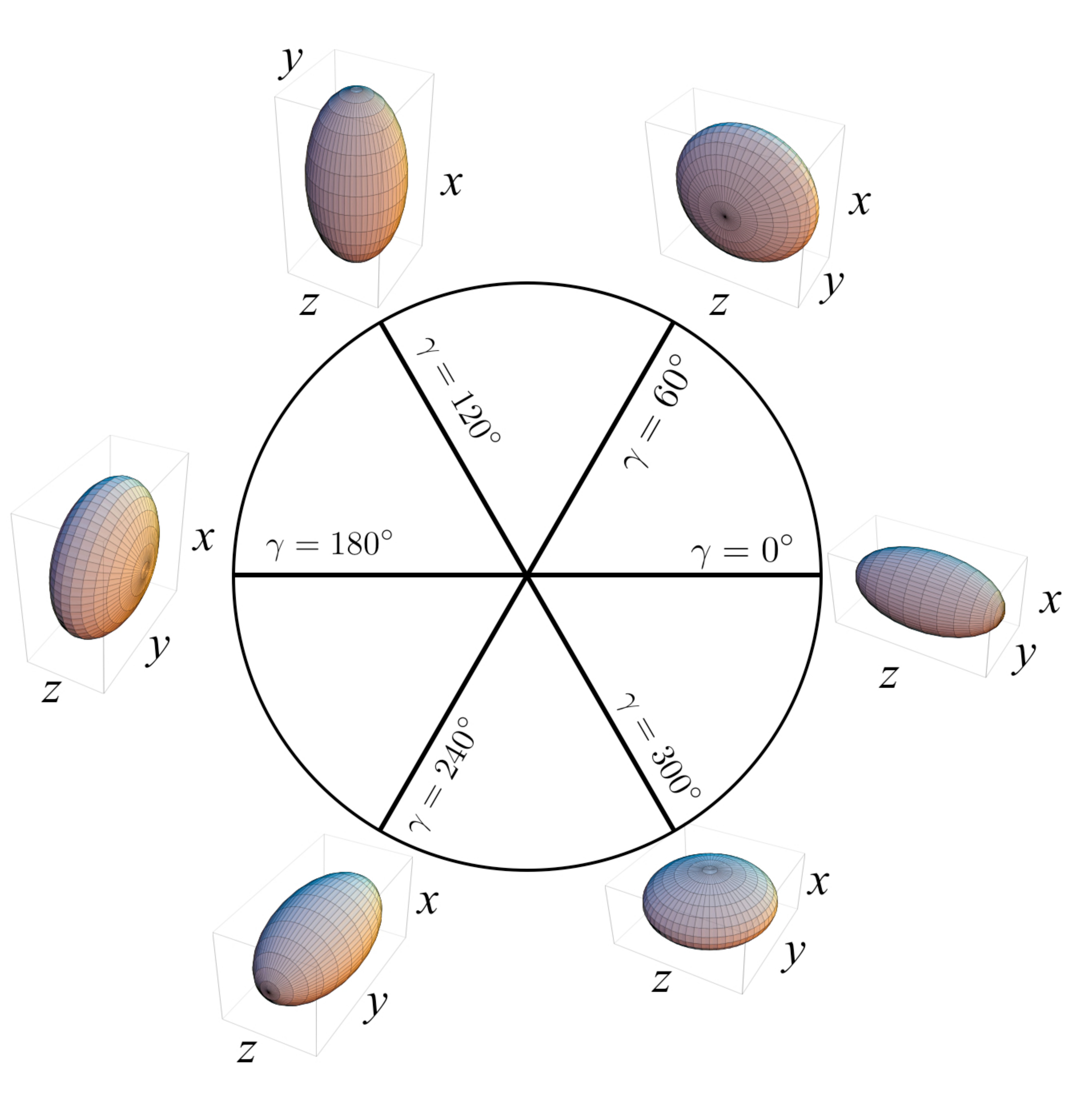}
\caption{(Color online) Orientations of the intrinsic deformation as a function of the $\gamma$ parameter. $\gamma=0^{\circ},120^{\circ},240^{\circ}$ and $\gamma=60^{\circ},180^{\circ},300^{\circ}$ correspond to axial symmetric prolate and oblate shapes, respectively.}
\label{Fig3}
\end{figure}

\begin{figure}[t]
\centering
\includegraphics[width=0.45\textwidth]{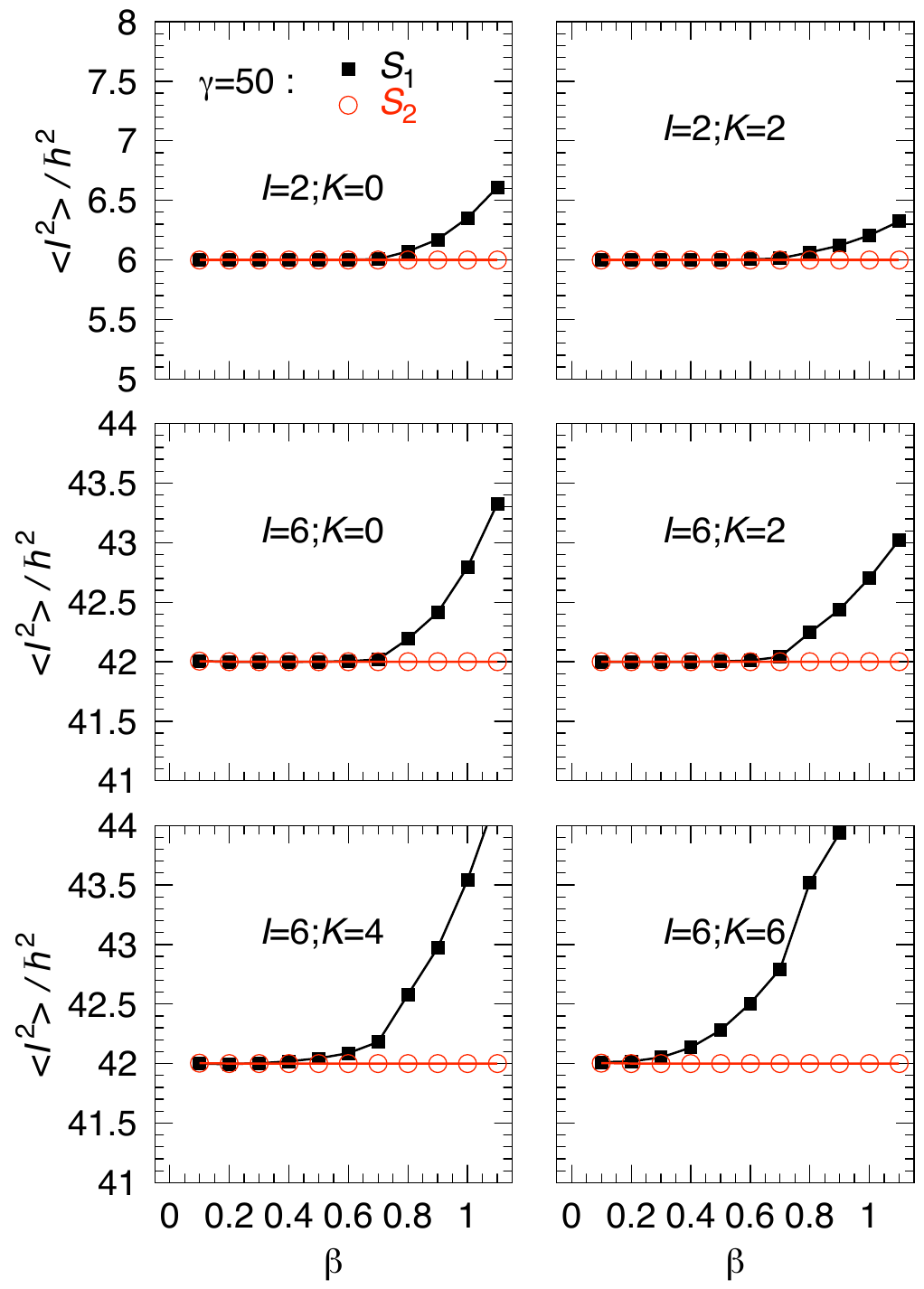}
\caption{(Color online) Expectation values of the total angular momentum operator calculated with angular momentum projected states $|IK\rangle$ as a function of the $\beta$ deformation ($\gamma=50^{\circ}$) and for different sets of integration points in the Euler angles $(a,b,c)$ (red circles $S_2=(16,16,32)$, black filled boxes $S_1=(6,16,12)$. The top and bottom panels correspond to $I=2$ and $I=6$, and their corresponding $K$ values, respectively. }
\label{Fig4}
\end{figure}

\begin{figure}[t]
\centering
\includegraphics[width=0.45\textwidth]{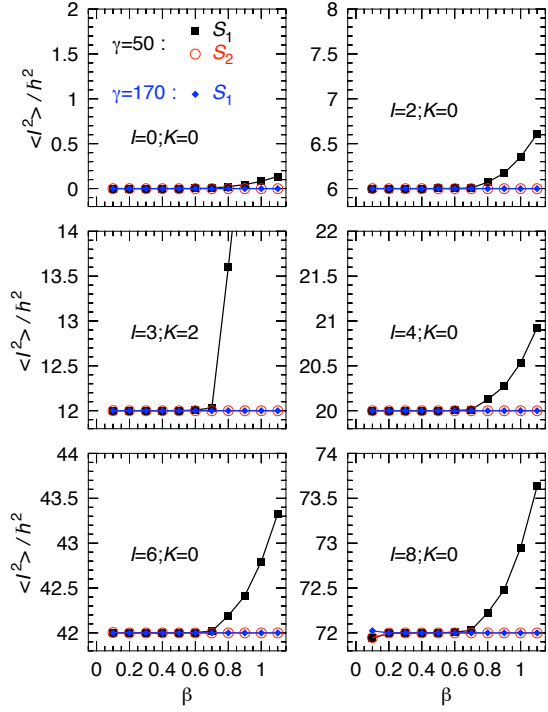}
\caption{(Color online) Expectation values of the total angular momentum operator between angular momentum projected states $I=0,2,4,6,8; K=0$ and $I=3;K=2$ as a function of the $\beta$ deformation and for different sets of integration points in the Euler angles and orientation of the intrinsic axes, red circles $S_2=(16,16,32)$, black filled boxes $S_1=(6,16,12)$ with $\gamma=50^{\circ}$ and blue filled diamonds with $\gamma=170^{\circ}$.}
\label{Fig5}
\end{figure}

\begin{figure}[t]
\centering
\includegraphics[width=0.7\textwidth]{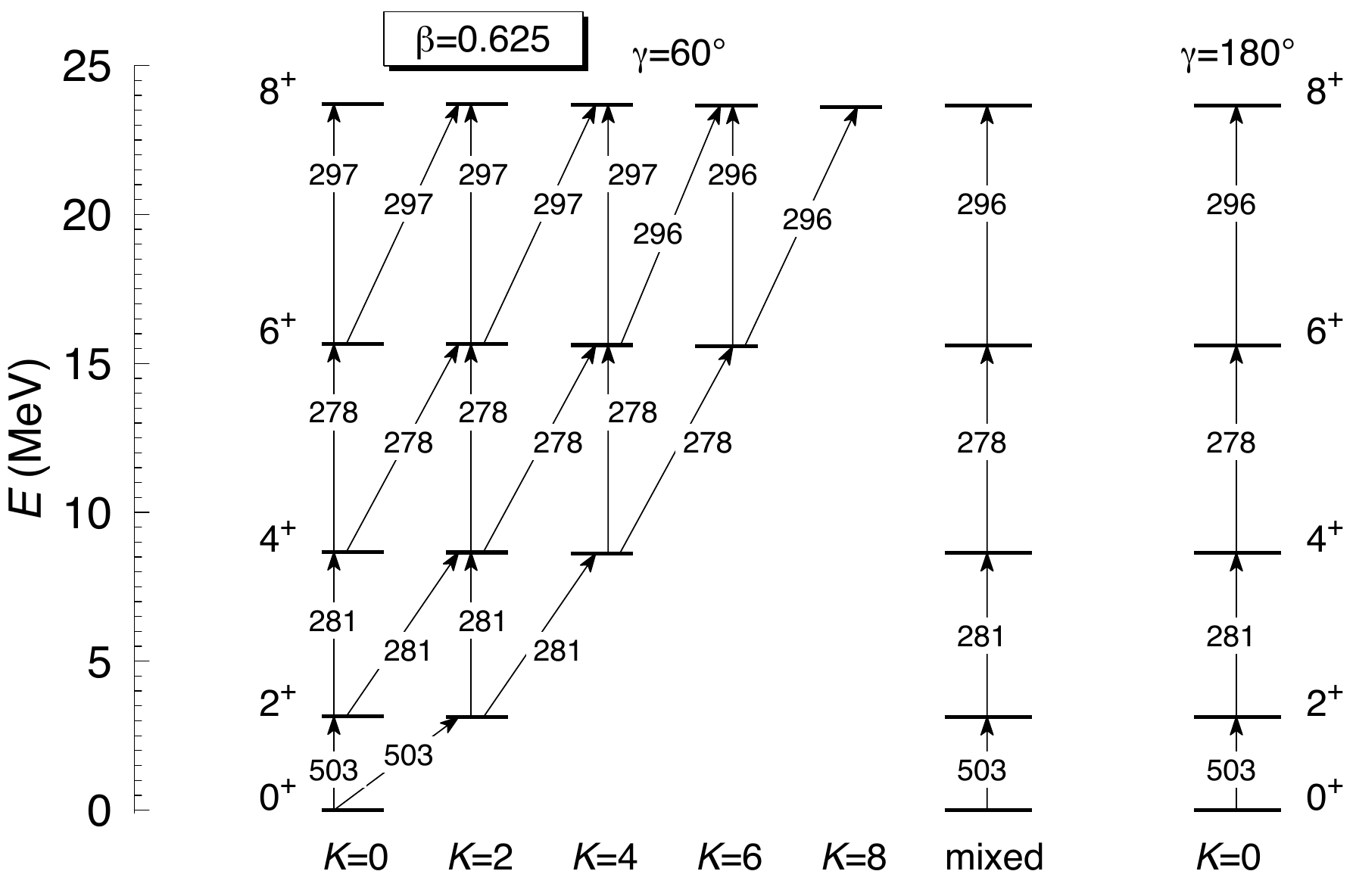}
\caption{(Left panel) excitation energies and B(E2) (in $e^{2}$fm$^{4}$ units) values for the states before and after $K$ mixing -$|IMK\rangle$ and $|IM\rangle$ respectively- with $\beta=0.625$ and $\gamma=60^{\circ}$. (Right Panel) excitation energies and BE(2) values for the state $|IMK=0\rangle$ with $\beta=0.625$ and $\gamma=180^{\circ}$}
\label{Fig6}
\end{figure}

\begin{figure}[t]
\centering
\includegraphics[width=0.7\textwidth]{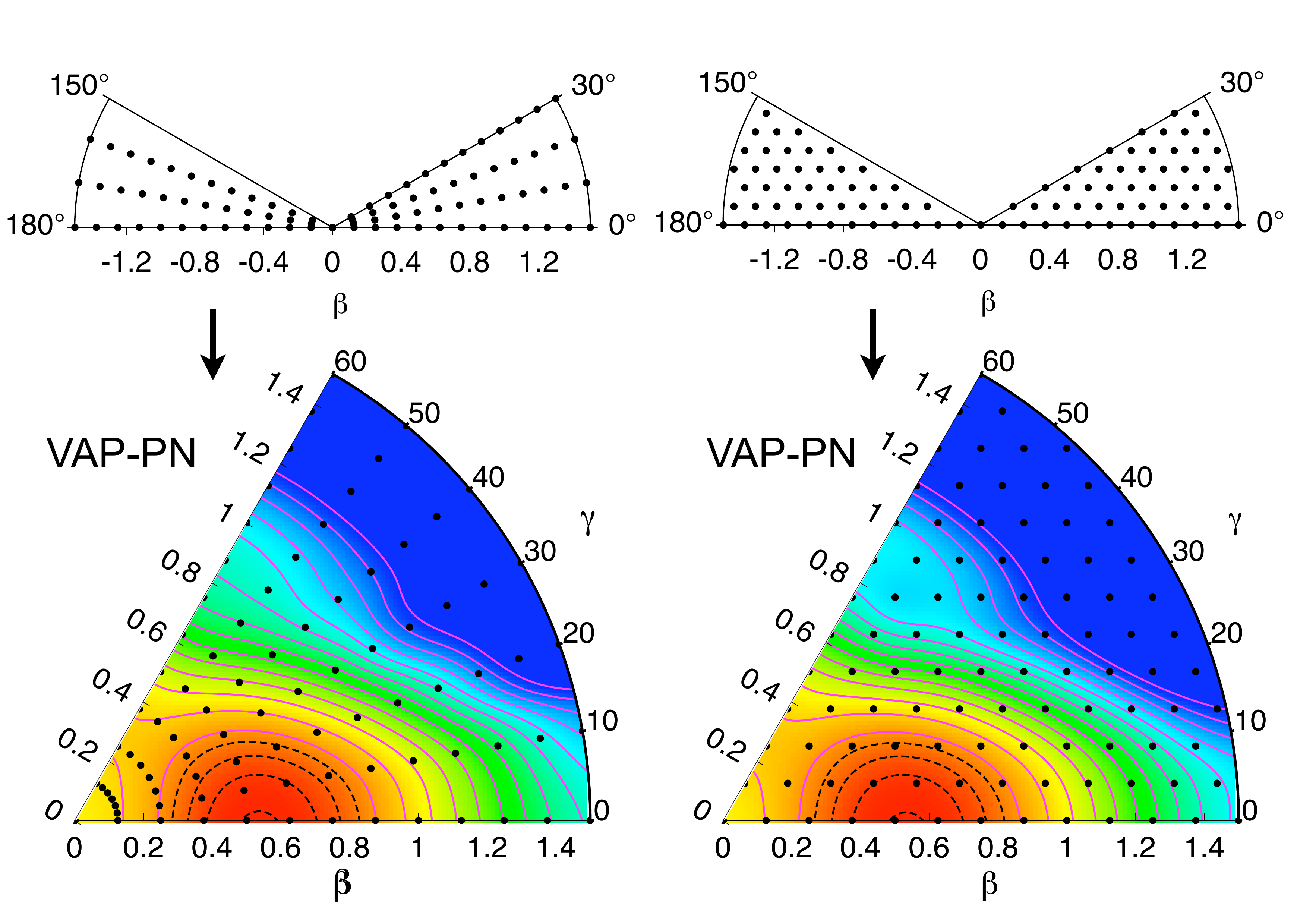}
\caption{(Color online) Mesh of points using constant step in $\beta$ and $\gamma$ (top left) or triangle division (top right) and the corresponding calculated VAP-PN potential energy surfaces (lower panels) transformed to the interval $\gamma$  $\epsilon [0^{\circ},60^{\circ}]$. The energy is normalized to the minimum of the PES ($-196.01$ MeV) and the contour lines are divided in 1 MeV (black dashed lines) and 2 MeV steps (continuous magenta lines).}
\label{Fig7}
\end{figure}

In this section we will study some aspects of the simultaneous particle number and angular momentum projection with triaxial shapes. Firstly, it is important to note that the parametrization of the quadrupole deformation in terms of $(\beta,\gamma)$ variables gives a triple degeneracy in the range $0^{\circ}\leq\gamma\leq 360^{\circ}$ if we consider time-reversal conserving wave functions \cite{RingSchuck}. This degeneracy corresponds to the three possible orientations of the intrinsic axis $I_{3}$ with respect to the $z-$axis (see Fig. \ref{Fig3}). Therefore, the interval $0^{\circ}\leq\gamma\leq 60^{\circ}$ covers all the possible quadrupole deformations. However, we can take advantage of this symmetry first to improve the convergence of the integral in the Euler angles that must be carried out in the PNAMP calculation (Eq. \ref{IMK_WF}) and second  to perform consistency checks of the results.
We now study the convergence of the integral in the Euler angles with respect to the number of integration points in $\Omega=(a,b,c)$. We have considered the symmetries of the intrinsic wave function reducing the integration interval to $(0\leq a\leq \pi/2,0\leq b\leq \pi/2,0\leq c\leq \pi )$ (see Refs. \cite{EnamiCHF_PPQ_99,BendHFB_SK_08,RingAMP_Rel_09}) and we have used Gaussian-Legendre quadratures for the numerical integration. As in the axial case, the number of integration points for the particle number projection is kept to $N_{Fom}=9$, which is sufficient to get eigenstates of the particle number operators. Naturally, the best candidate to check the convergence of the angular momentum projection is the expectation value of the total angular momentum operator $\hat{I}^{2}$ that, considering Eq. \ref{I2}, must be:
\begin{equation}
\langle\hat{I^{2}}\rangle_{IK}=\frac{\int\mathcal{D}^{I*}_{KK}(\Omega)\langle \Phi|\hat{I}^{2}\hat{R}(\Omega)P^{N}P^{Z}|\Phi\rangle d\Omega}{\int\mathcal{D}^{I*}_{KK}(\Omega)\langle \Phi|\hat{R}(\Omega)P^{N}P^{Z}|\Phi\rangle d\Omega}=\hbar^{2}I(I+1).
\end{equation}
The convergence in the number of integration points depends on three factors, namely the orientation of the intrinsic axes, the values of $(I,K)$ and the deformation $\beta$. Let us start with  the two latter factors. In Fig. \ref{Fig4} we plot  the mean value of the total angular momentum operator as a function of $\beta$ for projected wave functions with $I=2,6$ and a fixed value of $\gamma=50^{\circ}$. The integration has been performed with two sets of integration points in $(a,b,c)$, $S_{1}=(6,16,12)$ and $S_{2}=(16,16,32)$. Here, we can observe that for the set $S_{2}$  the correct result of the eigenvalue is obtained for all $\beta$ and $I,K$. However, the set $S_{1}$  fails both for large values of $\beta$ for all $I,K$ and also for smaller deformations with high $K=4,6$. The poor performance of this choice is clearly seen in the latter case where substantial deviations from the correct number are observed. Therefore, as a rule of thumb, the larger the values of $(I,K)$ and $\beta$  the more integration points are needed to have good results. The final choice will be the one that is able to provide converged results for all ($I,K,\beta,\gamma$) values.
Taking into account that the symmetry axis corresponds to pure  $K=0$ states, one may assume that close to the symmetry axis only
small $K$-components are present. We therefore examine the role of the orientation of the intrinsic axes in the PNAMP method.  First, we explore the convergence of the angular momentum projection using the property given in Fig. \ref{Fig3} and projecting symmetric states with the same value of $\beta$ but with $\gamma'=120^{\circ}+\gamma$.  If our assumption is right, we could reduce the number of integration points using   instead a given wave function an {\it equivalent} intrinsic wave function with an orientation closer to the $K=0$ case. In Fig. \ref{Fig5} we plot as a function of $\beta$ the expectation values of the angular momentum operator for intrinsic states with $\gamma=50^{\circ}$ and also with $\gamma'=170^{\circ}$. The sets of integration points are the same as in Fig. \ref{Fig4}. For the set  $S_{1}$ with $\gamma=50^{\circ}$  we observe again the loss of convergence whenever $\beta$ and $I$ increase. However, very much improved results are obtained for the same set of integration points, $S_{1}$, but projecting the wave functions with the  $\gamma=170^{\circ}$ orientation. In addition, the calculation with the set  $S_{2}$  reveals the numerical origin of the lack of convergence for the set $S_{1}$  with $\gamma=50^{\circ}$. Therefore, we will use this property to define the mesh in the $(\beta,\gamma)$ plane for performing GCM-PNAMP calculations as we will see below.\\
The analysis shown in Figs.~\ref{Fig4} and  \ref{Fig5} has been performed with diagonal matrix elements. Since in the GCM calculations we
have to consider also non-diagonal matrix elements, we have extended our study to this case. We find that in order to ensure a good convergence in all cases, the final set of integration points in the Euler angles has to be  chosen as $(N_{a}=8,N_{b}=16,N_{c}=16)$.
We can also exploit the degeneracy illustrated in Fig. \ref{Fig3} to perform a consistency test of the implementation of the PNAMP method \cite{BendHFB_SK_08}. Using symmetry properties of the point group $D_2$ it can be shown,  in the notation of eq.~\ref{IMK_WF}, that
\begin{equation}
\frac{\langle IMK;NZ;\beta\gamma=60^{\circ}| \hat{H}|IMK;NZ;\beta\gamma=60^{\circ}\rangle}{\langle IMK;NZ;\beta\gamma=60^{\circ}| IMK;NZ;\beta\gamma=60^{\circ}\rangle} = 
\frac{\langle I00;NZ;\beta\gamma=180^{\circ}| \hat{H}|I00;NZ;\beta\gamma=180^{\circ}\rangle}{\langle I00;NZ;\beta\gamma=180^{\circ}|I00;NZ;\beta\gamma=180^{\circ}\rangle},
\end{equation}
i.e., the projected energy calculated with a HFB wave function with $\gamma=60^{\circ}$ is $K$-independent and equal to 
the projected energy calculated with the HFB wave function with  $\gamma=180^{\circ}$. A similar relation applies for the transition probabilities.
In Fig. \ref{Fig6} we show the excitation energies and reduced transition probabilities B(E2) calculated  with the same oblate axially symmetric  wave function ($\beta=0.625$) but oriented differently in space with $\gamma=60^{\circ}$ (left panel) and $\gamma=180^{\circ}$ (right panel). 
As expected,  we find that the $\gamma=60^{\circ}$ excitation spectrum and transition probabilities are $K$-independent and therefore   identical to the mixed ones. A look to the right panel corroborates also that these quantities coincide with the ones generated with 
the   $\gamma=180^{\circ}$ intrinsic wave function.
Once we have analyzed the convergence and consistency of the PNAMP method for a given point in the $(\beta,\gamma)$ plane we can study the potential energy surfaces (PES) for the different approaches (VAP-PN, PNAMP with and without $K$-mixing). We explore first the role of the mesh of points needed to cover all different triaxial shapes. Given the better convergence properties for wave
functions  with a large $K=0$ component (compare Fig.~\ref{Fig5}), we divide the calculation into  two regions, $\gamma\in[0^{\circ},30^{\circ}]$ and $\gamma\in[150^{\circ},180^{\circ}]$ (see Fig.\ref{Fig7}). The last interval is equivalent to $\gamma\in[30^{\circ},60^{\circ}]$ and we will transform the results to it whenever we plot the different PES throughout this paper. Furthermore, the resolution of the PES is affected by the way we perform the discretization of the plane. In the lower panels of Fig. \ref{Fig7} we show the VAP-PN energy surfaces for a constant step division both in $\beta$ and $\gamma$ directions (left part) and for a division based on equilateral triangles (right part). The number of points is  $N_{points}=99$ in both cases. We observe that the distribution of the points in constant steps is not the best choice neither for small $\beta$, where for many points  almost degenerated states are obtained, nor for large $\beta$, where a loss of resolution in $\gamma$ is observed for increasing values of $\beta$. It is precisely in this region where the interpolation between distant points produces artifacts or wrong results in the PES such as spurious oscillations, as for example in the region  $(\beta\in[1.0,1.2], \gamma\in[20^{\circ},40^{\circ}])$ or softening of the contour plots $(\beta\in[0.6,1.1],\gamma\in[50^{\circ},60^{\circ}])$. This is rectified with a discretization based on triangles and the results presented hereafter are calculated with this mesh. Nevertheless, although only small differences around the minimum of the PES are obtained in the case of $^{24}$Mg, these effects will be enhanced for rather $\gamma$-soft and moderate $\beta$ deformed nuclei. In those cases, the division based on triangles will give much better results for the same number of total points included in the calculation and will save computing time with respect to the other mesh.
\section{Triaxial calculations  for $^{24}$Mg} \label{GCM-PNAMP}
In the previous sections we have studied several aspects needed to ensure a good performance of the full generator coordinate method with the particle number and triaxial angular momentum projected wave functions. This previous research is important because the full GCM-PNAMP calculation is very demanding in CPU-time and both convergence tests and the choice of the relevant parameters should be performed in advance, but nonetheless also checked afterwards. In this section the final results for $^{24}$Mg are presented, their calculation as mentioned above, have been done with the set of integration points in the Euler angles
 $(N_{a}=8,N_{b}=16,N_{c}=16)$.  We choose the triangular mesh with $N_{points}=99$ shown in Fig. \ref{Fig7} to solve the constrained particle number projection before the variation (VAP-PN) equations.  The intrinsic many body wave functions $|\Phi(\beta,\gamma)\rangle$ are expanded in a cartesian harmonic oscillator basis and the number of spherical shells included in this basis is $N_{shells}=7$ with an oscillator length of $b=1.01A^{1/6}$. In Fig. \ref{Fig7} the VAP-PN energy landscape is plotted showing a single and well defined minimum at $\beta=0.5, \gamma=0^{\circ}$ separated by $\sim 7.7$ MeV from the spherical point and $\sim 6.1$ MeV from the oblate saddle point at $\beta=0.25$. These results are consistent with the ones obtained in the axial calculation (see Fig. \ref{Fig1}) with the difference of having a saddle point in the $(\beta, \gamma)$ plane instead of a minimum on the oblate side. Similar PES are obtained for Skyrme (HFB with particle number projection after variation (PN-PAV) included) \cite{BendHFB_SK_08} and Relativistic (BCS without PNP) \cite{RingAMP_Rel_09} interactions although a softer surface between the spherical point and the minimum is obtained for the Skyrme interaction. \\
\begin{figure}[t]
\centering
\includegraphics[width=0.5\textwidth]{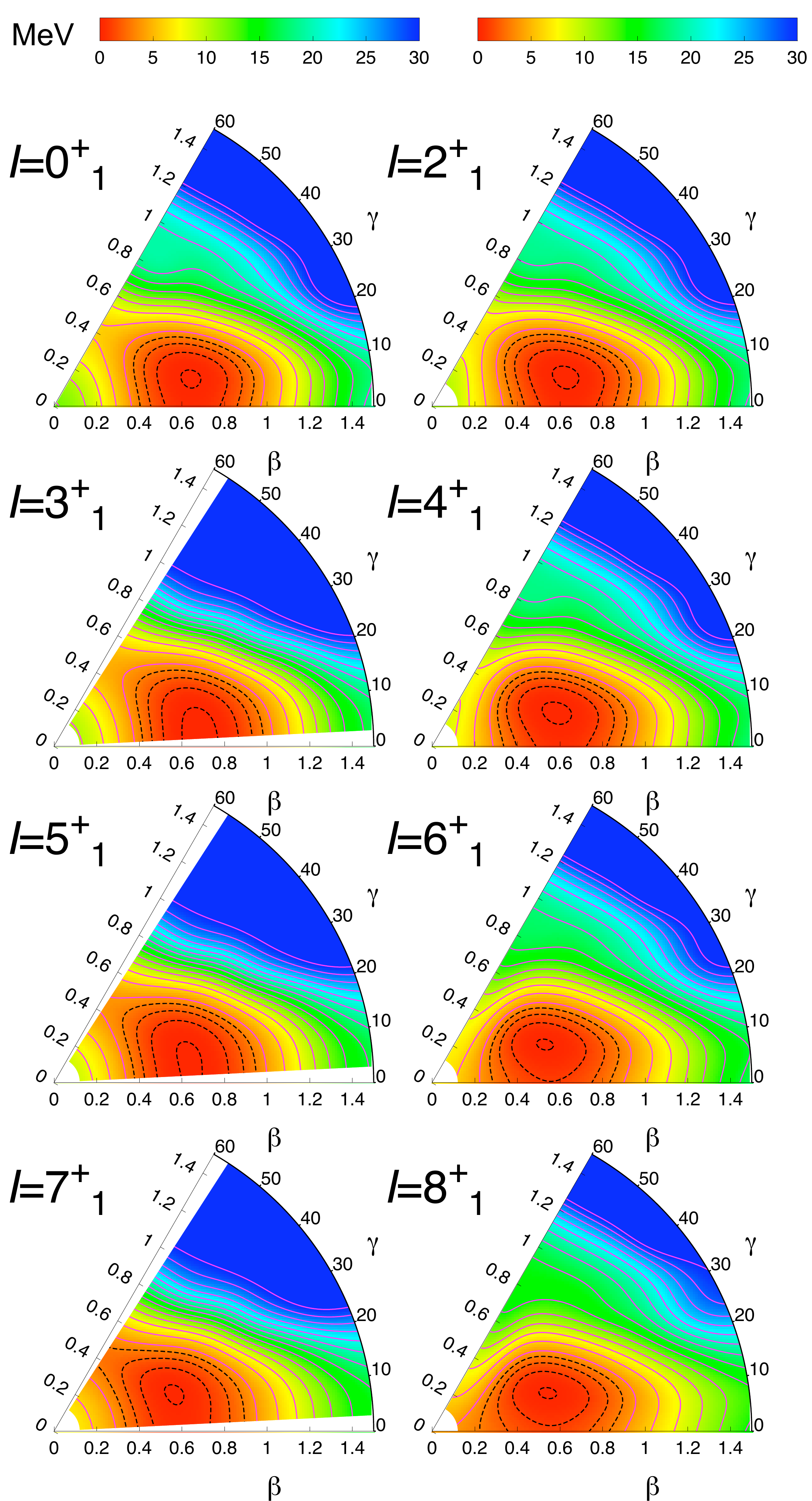}
\caption{(Color online) PNAMP potential energy surfaces including $K$-mixing in the $(\beta,\gamma)$ plane for $I=0-8$ and the first eigenvalues in $K$-space. The PES are normalized to the minimum of the surfaces (-200.74, -199.43, -194.04, -196.61, -190.86, -192.27, -186.09, -185.33  MeV for $I=0,2,3,4,5,6,7,8$, respectively). The contour lines are divided in 1 MeV (black dashed lines) and 2 MeV steps (continuous magenta lines). States with projected norm less than $10^{-6}$ are removed}
\label{Fig8}
\end{figure}
\subsection{Triaxial PNAMP potential energy surfaces and the RVAMPIR approach for $^{24}$Mg}
  The solution of the triaxial HWG equation, Eq.~\ref{HW_eq}, does not require to perform a separate angular momentum projection in the laboratory system for each component of the GCM basis states in the sense of Eq.~\ref{pnamp_state}. However, as in the axial case, we expect  the PNAMP potential energy surfaces to provide insight and a better interpretation of the configuration mixing calculations. We can also separate the energy gain due to the  triaxial AMP from the one due to the $(\beta,\gamma)$ configuration mixing. Furthermore, they are very important because the minima of these PES determine the associated  RVAMPIR solution. The PNAMP is an involved approach  that requieres the solution of the HWG equation, Eq.~\ref{HW_eq_pnamp}, to  include the $K$ mixing.  The HWG eigenstates, Eq.~\ref{pnamp_state}, provide real  eigenstates of the symmetry operators that can be used, as we shall see below, to generate energy spectra and to calculate transition probabilities.

 In Fig. \ref{Fig8} we plot the normalized PNAMP energy landscapes in the $(\beta,\gamma)$ plane for the lowest eigenvalue in the $K$-space for each angular momentum $I=0^{+}_{1}-8^{+}_{1}$ (see Eq.\ref{HW_eq_pnamp}). In addition, all the points close to the spherical one, and those close to axiality for odd values of $I$, have been removed for $I\neq0$ because their norm is very small. The first noticeable aspect is that the VAP-PN axial minimum of Fig. \ref{Fig7} becomes a saddle point, the minimum being displaced  towards larger $\beta$ values and $\gamma >  0^{\circ} $  for all values of the angular momentum, although the barriers between the
 new minima  and the axial prolate saddle points are less than 1 MeV. For $I=0^{+}_{1},2^{+}_{1}$ the minima are located in $(\beta\sim 0.7, \gamma\sim10^{\circ})$ while with increasing value of the angular momentum we observe a softening of the PES  and a displacement of the minimum to larger $\gamma$ and smaller $\beta$ deformation, $(\beta\sim0.65, \gamma\sim15^{\circ})$ for $I=4^{+}_{1},5^{+}_{1}$ and $(\beta\sim0.55, \gamma\sim17^{\circ})$ for $I=6^{+}_{1},7^{+}_{1},8^{+}_{1}$. We  also note that in the case of odd-$I$ values the softening of the PES  is in the $\gamma$ direction towards the oblate saddle point. The energy difference between the VAP-PN and the $I=0^{+}_{1}$ minima is $\sim4.6$ MeV while the gain in energy due to the inclusion of the triaxial degree of freedom, i.e, the difference between the triaxial minimum and the axial saddle point, is $\sim0.7$ MeV. Similar results have been reported with Skyrme and Relativistic interactions although these studies  of the  PNAMP-PES only extend to $I=0,2$ and the effect of increasing triaxiality with growing angular momentum has not been analyzed.\\
\begin{figure}[hbtp]
\centering
\includegraphics[width=\textwidth]{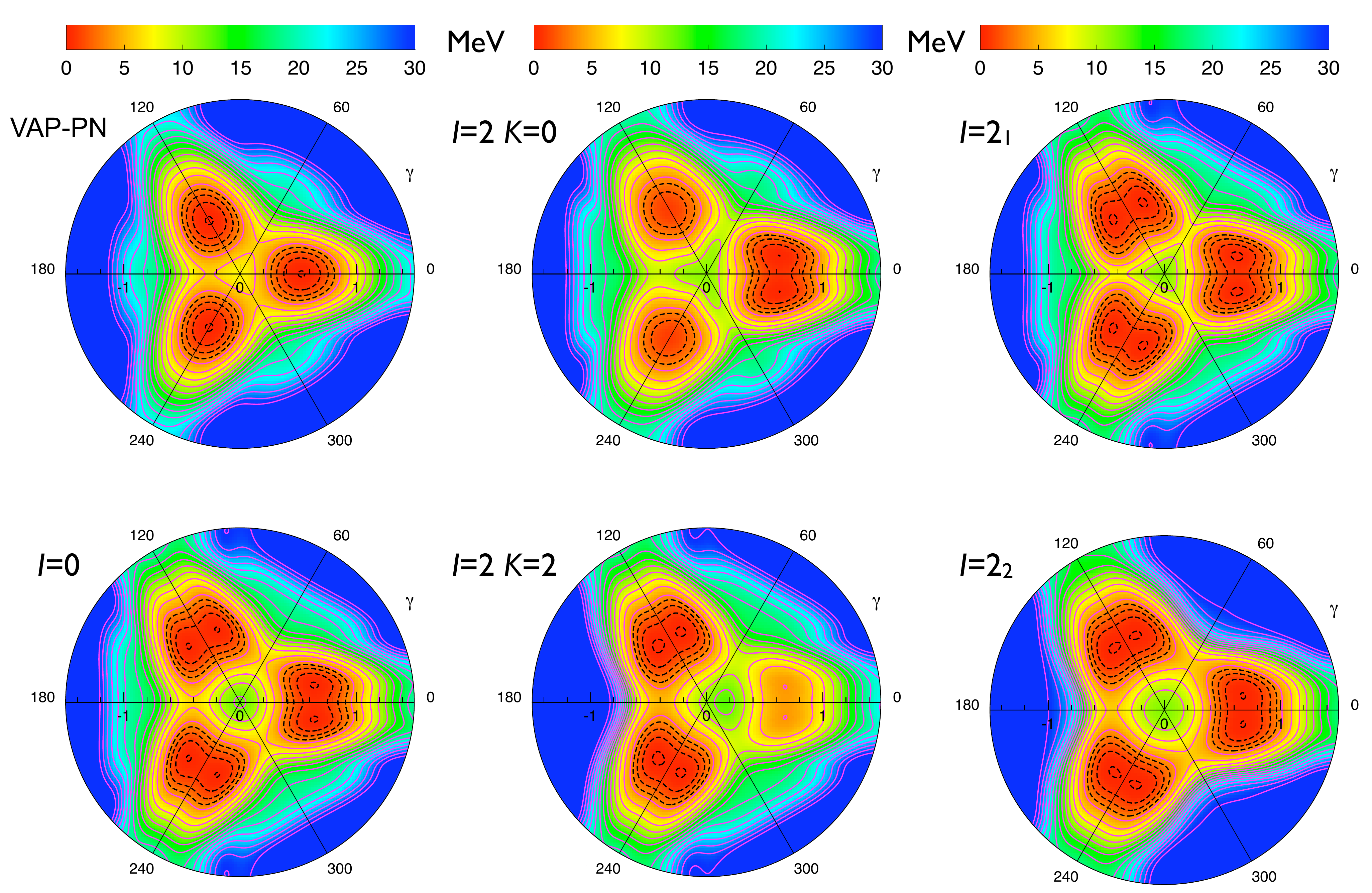}
\caption{ (Color online) Potential energy contour plots for $^{24}$Mg  in the $(\beta,\gamma)$ plane for   $\gamma= 0^{\circ} - 360^{\circ}$ in different approaches and angular momenta normalized to the corresponding minima. The contour lines are divided in 1 MeV (black dashed lines) and 2 MeV steps (continuous magenta lines). ({\bf Left panels}) Top:  Particle number projection (VAP), $E_{min}=-196.02$ MeV; bottom: PNAMP  approach for $I=0 \hbar$, $E_{min}=-200.74$ MeV. ({\bf Middle panels}) $IK$-projected energies according to Eq.~\ref{K-energy}. Top: $I=2, K=0$, $E_{min}=-199.42$ MeV; bottom: $I=2, K=2$, $E_{min}=-198.78$ MeV.  ({\bf Right panels}) Lowest eigenvalues of the PNAMP approach. Top: $I=2_1$, $E_{min}=-199.43$ MeV; bottom: $I=2_2$, $E_{min}=-195.18$ MeV.}
 \label{Fig9}
\end{figure}
\begin{table}[htdp]
\centering
\begin{tabular}{c|c|c|ccc}
\hline
\hline
$I^{+}_{\sigma}$&$(\beta,\gamma^{\circ})_{\rm min}$&$E_{\rm min} (MeV)$&$K=0$&$K=\pm2$&$K=\pm4$\\
\hline
\hline
$0^{+}_{1}$&(0.696, 8.95)&0.000&\bf{1.000}&---&---\\
$2^{+}_{1}$&(0.696, 8.95)&1.311&\bf{1.000}&0.000&---\\
$2^{+}_{2}$&(0.696, 8.95)&5.556&0.000&\bf{0.500}&---\\
$3^{+}_{1}$&(0.696, 8.95)&6.695&0.000&\bf{0.500}&---\\
$4^{+}_{1}$&(0.661, 19.1)& 4.129&\bf{1.000}&0.000&0.000\\
$4^{+}_{2}$&(0.661, 19.1)& 8.116&0.000&\bf{0.500}&0.000\\
$5^{+}_{1}$&(0.661, 19.1)&9.883&0.0000&\bf{0.499}&0.001\\
$6^{+}_{1}$&(0.545, 23.4)& 8.471&\bf{0.997}&0.001&0.000\\
$6^{+}_{2}$&(0.545, 23.4)&12.139&0.002&\bf{0.497}&0.002\\
$7^{+}_{1}$&(0.545, 23.4)&14.645&0.000&\bf{0.498}&0.002\\
$8^{+}_{1}$&(0.545, 23.4)&15.401&\bf{0.924}&0.036&0.002\\
\hline
\hline
\end{tabular}
\caption{$\beta$ and $\gamma$ coordinates of the triaxial PNAMP minima  after $K$-mixing as well as excitation energies and distribution of  $K$ components (i.e., $|H^{I;NZ;\sigma}_{K}(\beta,\gamma)|^{2}$ see Eq.~\ref{pnamp_state} and below) as a function of $I^\pi_{\sigma}$. The values of $(\beta,\gamma)_{\rm min}$  may not coincide exactly with those of Fig.~\ref{Fig8} because of the finite size of the grid used in the calculations. The quoted values are the actual ones used in the $K$-mixing calculation.
The $K=\pm 6, \pm 8$ components, not shown, are exactly zero. }
\label{table0}
\end{table}
   For an interpretation of  the configuration mixing calculations  it has become customary to plot  the diagonal matrix elements of the normalized Hamiltonian overlap, see Eq.~\ref{gcm_overlaps}, i.e., the $IK$-projected energy  
\begin{equation}
\mathcal{E}^{I;NZ}_K(\beta\gamma) =  \frac{\mathcal{H}^{I;NZ}_{K\beta\gamma K\beta\gamma}}
{\mathcal{N}^{I;NZ}_{K\beta\gamma K\beta\gamma}},
\label{K-energy}
\end{equation}
in the $(\beta,\gamma)$ plane for the different $K$-values.  Since the states $|IMK;NZ;\beta \gamma \rangle$ are not eigenstates
of the angular momentum in the laboratory frame their energies do not have a physical meaning. Furthermore the  $IK$-projected energy PES and wave functions depend on the orientation of the axis in Fig.~\ref{Fig3}.  To illustrate this point  we present in            Fig.~\ref{Fig9}  PES's calculated in three approaches  for different orientations of the nucleus  according to Fig.~\ref{Fig3}. We observe 
that, as expected,  one sixth of the circle (for instance $\gamma=0^\circ-60^\circ$)  is enough to describe the PES corresponding to the VAP-PN and the PNAMP ones (corresponding to $I=0_1$ and $I=2_1, 2_2$).  However  for the $K$-projected PES's ($I=2, K=0$ and
$I=2, K=2$) a semicircle  (for instance $\gamma=0^\circ-180^\circ$) is needed. Since in the laboratory system  all the six sectors 
are equivalent we explicitly see that it is the same  to use the region of $\gamma=150^\circ-180^\circ$ than 
$\gamma=30^\circ-60^\circ$, as we have done in the GCM calculations. The contour plots in the $IK$ projection can be easily understood looking at Fig.~\ref{Fig3}. For $I=2, K=0$, the collective AM is perpendicular to the z-axis and since semi-classically a rotor 
will prefer to rotate around the axis with the largest moment of inertia it is obvious that the energy minima are around $\gamma=0^\circ$.
For  $I=2, K=2$, the collective AM is parallel  to the z-axis and in this case the minima will be around $\gamma=120^\circ$ and $\gamma=240^\circ$.    Specially for the latter case we see that it can be dangerous to make interpretations based on the $\gamma=0^\circ-60^\circ$ sector. For nuclei with more mixing one should also care about the interpretation of the  $I=2, K=0$ surface.
 In any case, it is important to note that the $K$ value is not a good quantum number in the laboratory frame and therefore it is not an observable. In addition, the distribution of $K$ and the corresponding PES can change depending on the orientation of the intrinsic wave function (see Fig. \ref{Fig3}). Nevertheless,  in cases where the $K$-mixing is not very large this quantum number can be useful to give an interpretation of the different bands that could appear in the spectrum. As we will see below, $^{24}$Mg is a very good example of rather pure $|K|$ bands.   One should be aware, however, that even with rather pure $|K|=2$ bands, a mixing of $K=2$ and $K=-2$ takes place and since these states are not orthonormal pitfalls may appear.\\
As discussed in Sect.~\ref{RVAMPIR}, the minima of the PNAMP potential energy  surfaces  provide an approximation to an angular momentum projection in a variation after projection approach, which we have called RVAMPIR. 
   In  Table~\ref{table0} we present the  $(\beta,\gamma)$  of the minima  of the two lowest eigenstates  together with  the $K$-distribution of the corresponding wave functions.  As we observe there is almost no mixing,  the $0^+_1, 2^+_1, 4^+_1, 6^+_1, 8^+_1$ states are $K=0$ and the $ 2^+_2, 3^+_1, 4^+_2, 5^+_1, 7^+_1$,  $K=2$. Only at the highest angular momentum we observe very small $K$-mixing.
\begin{figure}[t]
\centering
\includegraphics[width=0.3\textwidth]{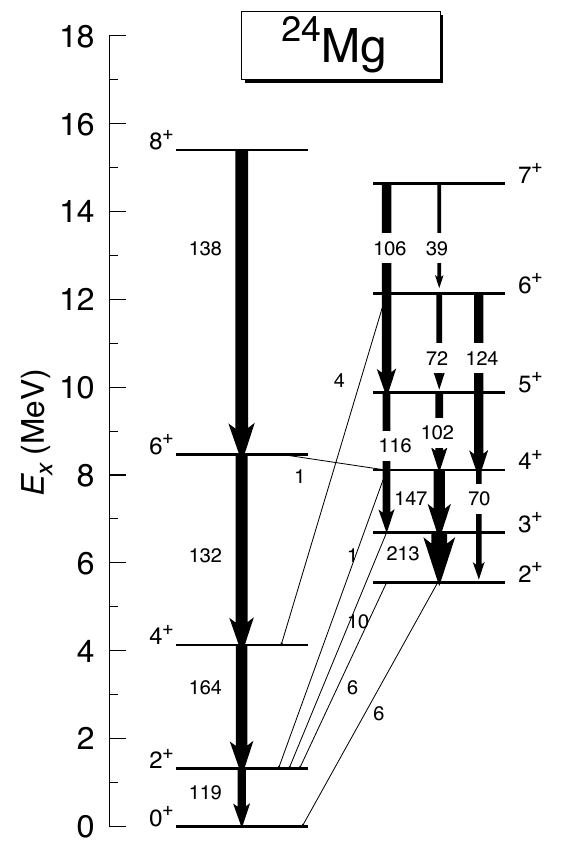}
\caption{ RVAMPIR spectrum and transition probabilities.}
\label{Fig10}
\end{figure}
  This is not the general rule. The amount of K-mixing depends strongly on the nucleus and on the $(\beta,\gamma)$ point.  As mentioned $^{24}$Mg seems to be a nucleus with rather small $K$-mixing.   
 The solution of the HWG equation, Eq.~ \ref{HW_eq_pnamp}, at the point  $(\beta^I_{\rm min},\gamma^I_{\rm min})$  provides  the RVAMPIR energies and  wave functions of the corresponding states.  In Fig.\ref{Fig10} we present the energy spectrum and the  calculated  transition probabilities.  Though we will discuss this figure in relation with the full GCM results we can compare with
the Axial PNAMP-GCM excitation spectrum of Fig.\ref{Fig2}.  The clear difference  is the presence of a well developed gamma band
in the RVAMPIR calculations. 
\begin{figure}[t]
\centering
\includegraphics[width=0.5\textwidth]{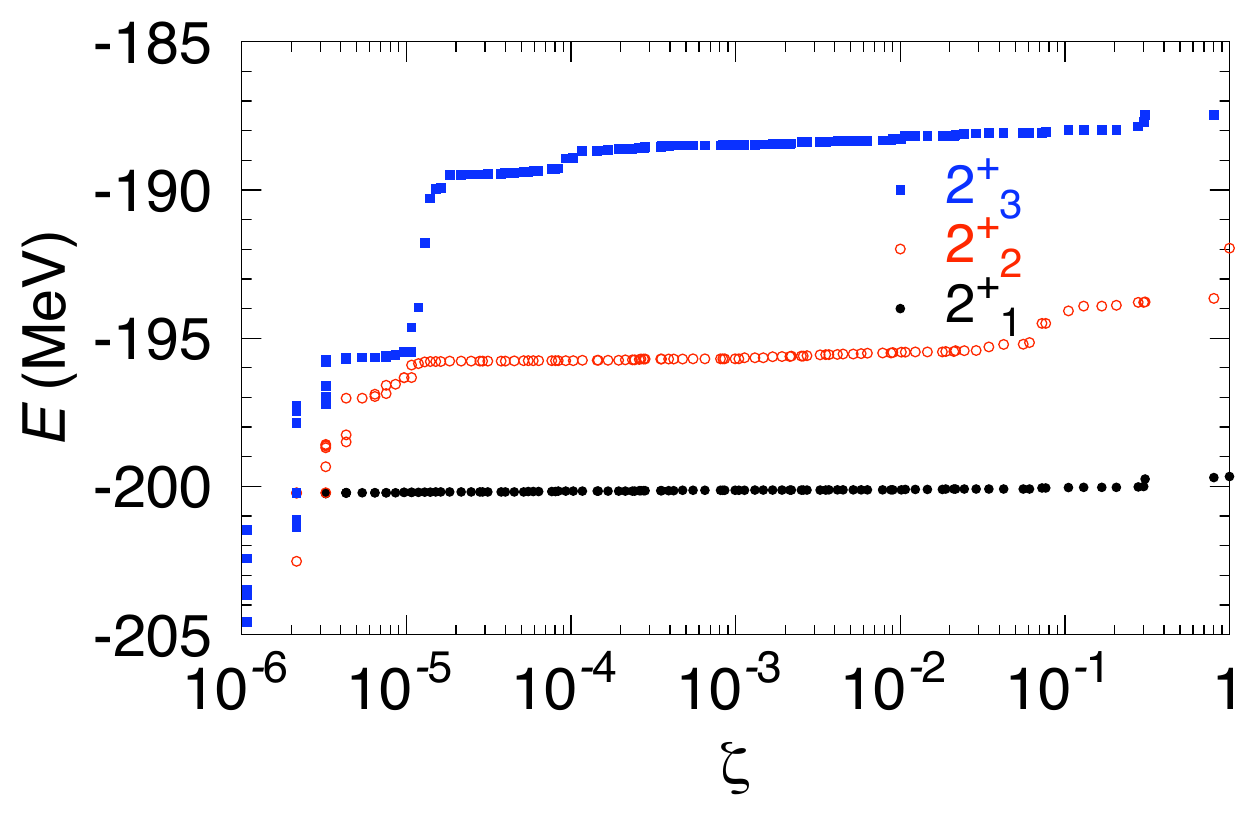}
\caption{(Color online) GCM-PNAMP energies ($I=2$) as a function of the corresponding norm eigenvalue, normalized to the highest value, used as cutoff in the definition of the natural basis (Eq. \ref{natstates}).}
\label{Fig11}
\end{figure}
\subsection{The configuration mixing calculations for $^{24}$Mg} 
The final step in the calculation of the spectrum is the GCM-PNAMP method, in which simultaneous mixing of the different deformations $(\beta,\gamma)$ and $K$ components is performed (see Eq. \ref{gcm_state}). As we mentioned in Sec. \ref{theo_frame}, we have to solve the HWG equations separately  for each value of the angular momentum. These generalized eigenvalue problems are solved removing the linear dependence of the states with the definition of the orthonormal natural basis (Eq. \ref{natstates}). In order to avoid spurious states in this basis, we have to define a cutoff parameter, $\zeta$, to determine the states in the natural basis, see Eq.~\ref{natstates} and text below. The convergence of the triaxial PNAMP-GCM method is showed in Fig. \ref{Fig11} where the  lowest three energy values obtained for $I=2$ are represented as a function of the parameter $\zeta$. Here we distinguish a region of large $\zeta$ in which the energies are decreasing followed by a range of values where the energies are nearly constant. The appearance of these \textit{plateaus} is the signature of the convergence of the GCM method \cite{Bonche_NPA_90}. We observe that this \textit{plateau} is better defined  for the  $2^{+}_{1}$ and  $2^{+}_{2}$ states as compared to the  $2^{+}_{3}$. Finally, for small values of $\zeta$ the linear dependence shows up and we obtain meaningless values for the energy. The final choice for $\zeta$ is a value around which we observe a wide \textit{plateau} for all the levels of interest. This value must be kept constant  for a given angular momentum in order to guarantee the orthogonality of the corresponding wave 
functions. This analysis has been performed for different values of the angular momentum showing in all cases a behavior similar to the one presented in Fig. \ref{Fig11}. Eventually, we have chosen $\zeta=10^{-3}$ as the final value, similar to the one found in 
Ref.~\cite{RingGCM_Rel_10}.  This procedure can be complemented by  an inspection  of the shape of the collective wave
function as a function of $\zeta$. \\
 \begin{figure}[t]
\centering
\includegraphics[width=\textwidth]{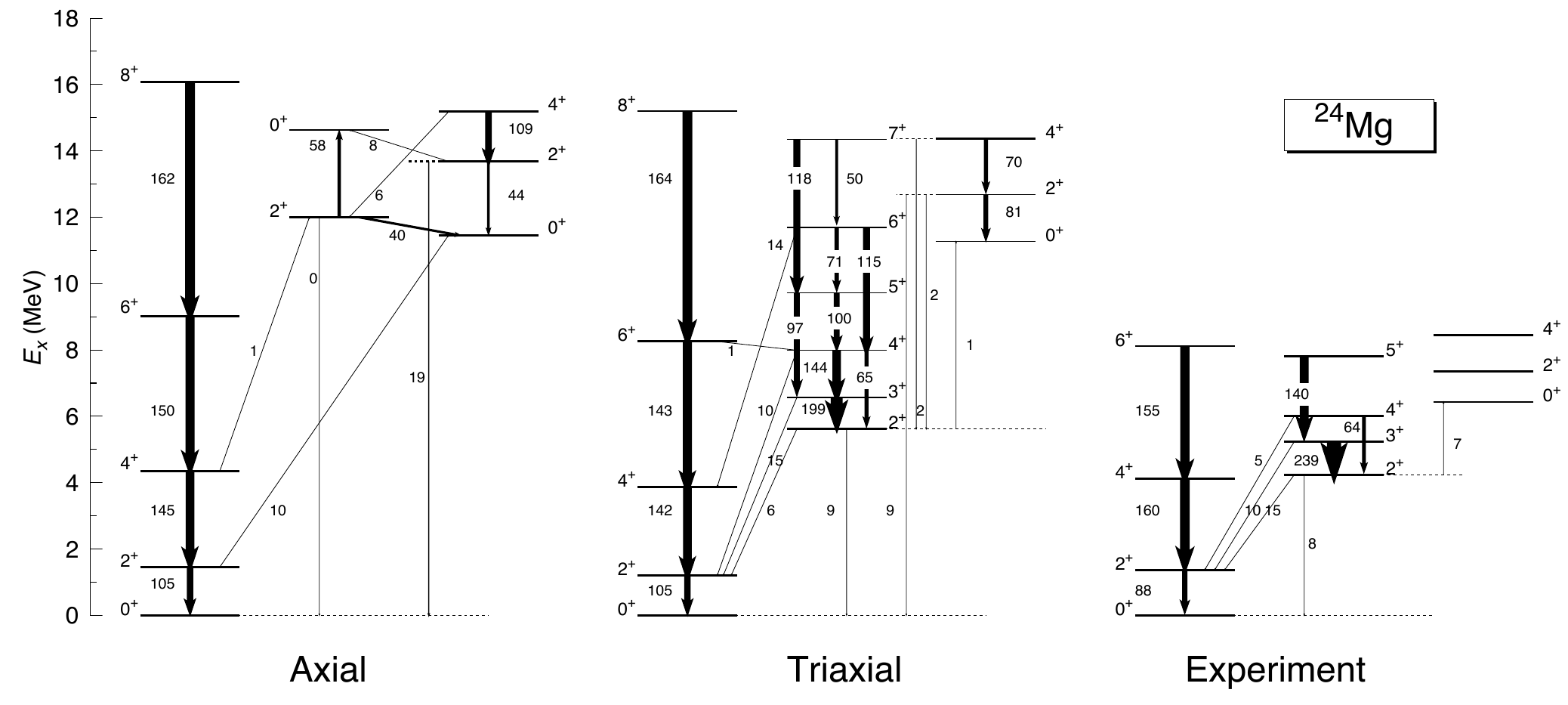}
\caption{Calculated  excitation energies and reduced transition probabilities B(E2) (in $e^2 fm^4$)  in $^{24}$Mg obtained using axially symmetric (left) and  triaxial (middle) GCM-PNAMP approaches compared to the  experimental values (right). The widths of the arrows are proportional to the corresponding B values . The experimental values are taken from \cite{Data}}
\label{Fig12}
\end{figure}
 In  the central panel of Fig. \ref{Fig12} we now  plot the spectrum of $^{24}$Mg extracted from the triaxial GCM calculations. We classify the different levels in three bands according to the corresponding B(E2) values. The ground state band is formed by a sequence of even  angular momentum states with a level spacing very similar to that of a rotational band. The first excited band consists of  states with $I=2,3,4,5,...$ as  expected for a $\gamma$ band. The third band is built of even-$I$ states on top of the second $0^{+}_{2}$ state.
We can also compare the absolute value of the ground state energy calculated with different approaches. Evidently, the lowest value is obtained with the triaxial GCM-PNAMP method (-201.36 MeV) while -200.74 MeV and -200.67 MeV are the results for RVAMPIR and axial GCM-PNAMP approximations respectively. Comparing the first two values we  observe that the energy gained by mixing different shapes is $\approx$0.5 MeV, much less than the correlation provided by PN and/or AM restoration. On the other hand, the inclusion of the triaxial degree of freedom within the GCM framework gives a similar energy gain ($\approx$0.5 MeV) due to the fact that the ground state -and also the whole band built on top of it- is already well described by an axial calculation in this particular nucleus. Major changes, as we will see below, are however found for the excited bands.
\begin{table}[htdp]
\centering
\begin{tabular}{c|c|ccccc}
\hline
\hline
$I^{+}_{\sigma}$&$E$(MeV)&$K=0$&$K=\pm2$&$K=\pm4$&$K=\pm6$&$K=\pm8$\\
\hline
\hline
$0^{+}_{1}$&0.000&\textbf{1.000}&---&---&---&---\\
$2^{+}_{1}$&1.202&\textbf{0.922}&0.039&---&---&---\\
$4^{+}_{1}$&3.875&\textbf{0.904}&0.016&0.032&---&---\\
$6^{+}_{1}$&8.256&\textbf{0.882}&0.011&0.015&0.033&---\\
$8^{+}_{1}$&15.198&\textbf{0.834}&0.040& 0.017&0.016&0.010\\
\hline
\hline
$2^{+}_{2}$&5.616&0.188&\textbf{0.406}&---&---&---\\
$3^{+}_{1}$&6.564&0.000&\textbf{0.500}&---&---&---\\
$4^{+}_{2}$&7.990&0.081&\textbf{0.413}&0.046&---&---\\
$5^{+}_{1}$&9.718&0.000&\textbf{0.407}&0.093&---&---\\
$6^{+}_{2}$&11.688&0.078&\textbf{0.398}&0.007&0.056&---\\
$7^{+}_{1}$&14.349&0.000&\textbf{0.379}&0.036&0.085&---\\
\hline
\hline
$0^{+}_{2}$&11.265&\textbf{1.000}&---&---&---&---\\
$2^{+}_{3}$&12.686&\textbf{0.958}&0.021&---&---&---\\
$4^{+}_{3}$&14.363&\textbf{0.795}&0.048&0.055&---&---\\
\hline
\hline
\end{tabular}
\caption{Decomposition  of the norm of GCM-PNAMP collective wave functions into $K$ components ($\sum_{\beta,\gamma}|F^{I;NZ;\sigma}_{K}(\beta,\gamma)|^{2}$) for the first, second and third bands. The highest values are printed in boldface. The excitation
energies of the corresponding states are also provided.}
\label{table1}
\end{table}
Concerning the transition probabilities, we observe strong electric quadrupole intraband transitions while the interband E2 transitions are much weaker. This fact indicates  different underlying structures of the bands  and the absence of mixing between them. We can study the nature of these bands decomposing the collective wave functions $|F^{I;NZ;\sigma}(\beta,\gamma)|^{2}$ (Eq. \ref{coll_wf}) into their $K$ components, $|F^{I;NZ;\sigma}_{K}(\beta,\gamma)|^{2}$, summing the contribution of all deformations $(\beta,\gamma)$ for each $K$. The result is shown in Table \ref{table1} , where we clearly  observe that the first and third bands are rather pure $K=0$  while the second band corresponds mainly to $|K|=2$ states. Furthermore, we see that for each level the $\pm K$ components have the same values, as a direct consequence of the time-reversal conservation of the intrinsic wave functions. \\
 The distribution of the states within these bands is supported by the values of the spectroscopic quadrupole moments:
\begin{eqnarray}
Q(I\sigma)=\sqrt{\frac{16\pi}{5}}\left(\begin{array}{ccc}
I & I & I \\
I & 0 & -I\end{array}\right)\langle I;NZ\sigma||\hat{M}^{\mathrm{elec}}_{2}||I;NZ\sigma\rangle
\end{eqnarray}
where $\hat{M}^{\mathrm{elec}}_{2\mu}=er^{2}Y_{2\mu}(\theta,\phi)$ are the electrical quadrupole moment operators. In the collective rotational model the spectroscopic quadrupole moments for a given $|K|$-band take the simple form \cite{RingSchuck}:
\begin{equation}
Q_{coll}(I,K)=Q_{0}\frac{3K^{2}-I(I+1)}{(I+1)(2I+3)}
\end{equation}
with $Q_{0}$ a constant deformation of the intrinsic macroscopic state. In Fig. \ref{Fig13} we compare the triaxial results with the values given for the collective rotational model with $K=0$ and $K=2$ -normalized to $I=2$. Here we can  clearly observe that the ground state band corresponds to a rotational band ($K=0$) while the second band matches to a $\gamma$-band ($K=2$) and the third band cannot be described in this simple picture.\\ 
\begin{figure}[t]
\centering
\includegraphics[width=0.5\textwidth]{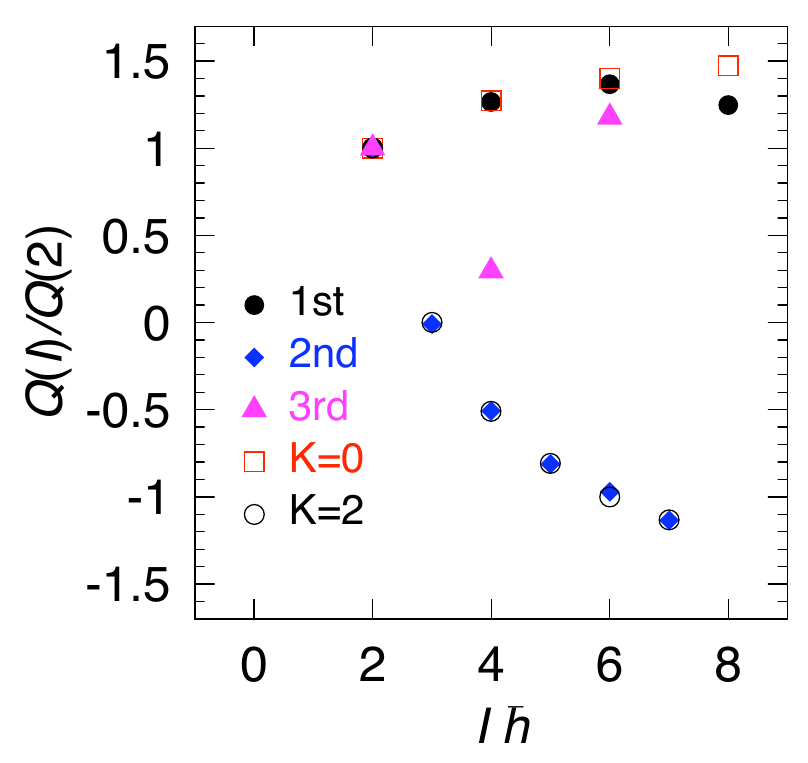}
\caption{(Color online) Spectroscopic quadrupole moments as a function of the angular momentum $I$ calculated for the states in the first (black bullets), second (blue diamonds) and third (magenta triangles) bands and for the vibrational-rotational collective model with $K=0$ (red boxes) and $K=2$ (black circles).}
\label{Fig13}
\end{figure}
We also plot the  probability distribution $|F^{I;NZ;\sigma}(\beta,\gamma)|^{2}$ of each GCM state in the $(\beta,\gamma)$ surface (Fig. \ref{Fig14}) summing all the possible $K$ components. The most noticeable aspect is that all the states belonging to the same band have a very similar  probability distribution in the $(\beta,\gamma)$ plane and that the overlap  between states of different bands is small.  One could assume that
these facts will lead to the intraband and interband B(E2) values shown in Fig. \ref{Fig12}.  However, as we shall see below, the reason for the small interband transitions seems to be more related to a $K$-hindrance aspect based on the fact that the ground band is a pure
$K=0$  and the $\gamma$-band a rather pure $|K|=2$ band.  In particular, all the states in the first band have a well defined maximum at $(\beta\sim0.58, \gamma=0^{\circ})$ and the probability drops rather symmetrically in the $\beta$ and $\gamma$ directions.
 For the second band, the probability distribution is concentrated in a region of the plane with $(\beta\in[0.4-1.0], \gamma\in[0^{\circ},35^{\circ}])$ with maxima around $(\beta\sim0.7,\gamma\sim18^{\circ})$. Finally, the states belonging to the third band show a high probability of having spherical shape $(0^{+}_{2})$ or slightly prolate  $(2^{+}_{3},4^{+}_{3},6^{+}_{3})$ 
 combined with a non-negligible mixing of strongly deformed states in the range of $\beta\in[0.8,1.3],\gamma\in[0^{\circ},30^{\circ}]$.
   The  PNAMP-PES of  Fig.~\ref{Fig8}  can help to understand the probability distribution  of the HWG equation. Looking at the ground band panels  $(0^{+}_{1},2^{+}_{1},4^{+}_{1},6^{+}_{1},8^{+}_{1})$ of Fig.~\ref{Fig8}  we find that all show soft triaxial minima close to the axial axis, the contour lines being  elongated along the radial direction and rather steep along the $\gamma$ angle. These states will mix with
 the mirrored ones at $\gamma=0^\circ$, see Fig.~\ref{Fig9}, and as a result distributions with a peak at $\gamma=0^\circ$ similar to the
 ones in the left panels of Fig.~\ref{Fig14} are expected.   If we now concentrate on the panels $(3^{+}_{1},5^{+}_{1},7^{+}_{1})$,  representative of the $\gamma$ band, we observe contour lines centered around a soft slightly triaxial minimum. These contours, at variance with the ones of the ground band, are softer in the $\gamma$ angle.  We found that the members of the $\gamma$ band are
 rather pure $K=2$ states,  that means, the norms of the states along the symmetry axis $(\gamma=0)$ are zero.  This axis acts  as
 a barrier between the states above  $\gamma=0^{\circ}$ and the mirrored ones hindering the mechanism described for the ground band. As a result  distributions similar to the ones in the middle panels of Fig.~\ref{Fig14} will be obtained. \\
\begin{figure}[t]
\centering
\includegraphics[width=0.5\textwidth]{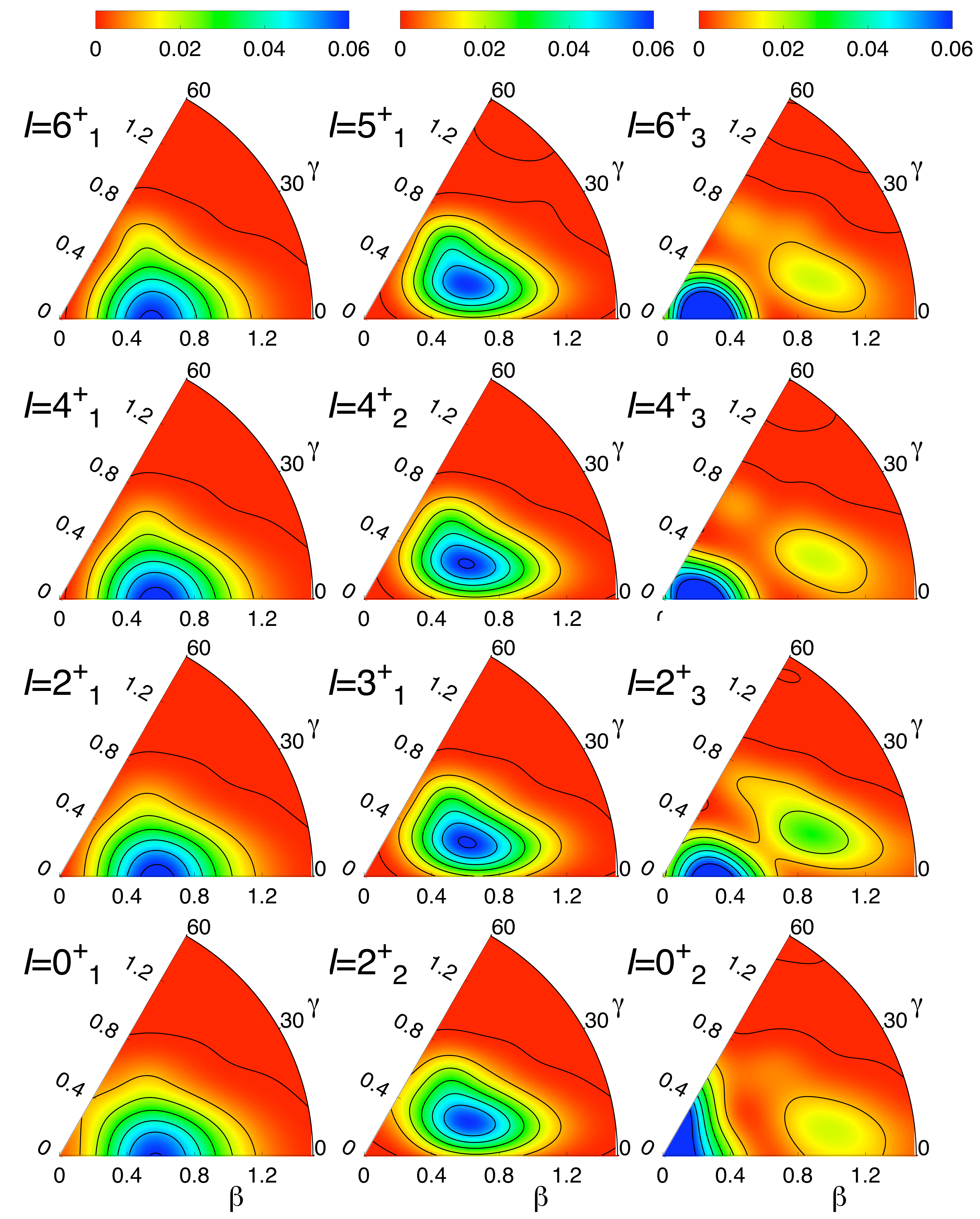}
\caption{(Color online) GCM-PNAMP collective wave functions $|F^{I;NZ;\sigma}(\beta,\gamma)|^{2}$ for the ground state (left), second (middle) and third (right) bands, respectively. Contour lines are separated in 0.01 units.}
\label{Fig14}
\end{figure}
In Fig. \ref{Fig12} we have also compared the triaxial results with axial calculations. In order to  better understand the results of this comparison, we  investigate first the relationship between the axial and triaxial collective wave functions. The axial states emerge from the $\gamma=0^{\circ}-180^{\circ}$ path of the $K=0$ component of the corresponding triaxial states. In particular, we can relate the ground state bands in both approaches and also the axial $0^{+}_{2},2^{+}_{3},4^{+}_{2}$ with the triaxial $0^{+}_{2},2^{+}_{3},4^{+}_{3}$ states (see Figs. \ref{Fig1} and \ref{Fig14}). Hence, the comparison between the triaxial and axial calculations reveals that both the energies and reduced transition probabilities of the ground state band are very similar in both cases, as expected. Nevertheless, the small $K$-mixing for $I\neq0$ lowers the excitation energies of higher angular momentum and therefore, the triaxial ground state band is slightly compressed with respect to the axial band. This effect, although small, helps to improve the description of the moments of inertia within the GCM-PNAMP framework. Larger differences between the axial and triaxial calculations appear for the second and third bands. Obviously, the axial calculations are unable to describe the $\gamma$-band but also the energies and B(E2) for the third triaxial band ($K=0$ mainly) are modified with respect to the corresponding ones in the axial case. This difference is due to both the small $K$-mixing and  the triaxial configuration around $\beta\sim1.0$ that appears already for $I=0$ (see Fig. \ref{Fig14}). \\  
In Table~\ref{table2} we present the average intrinsic deformation parameters and the spectroscopic quadrupole moments obtained for  the axial and triaxial calculations. In general the average $\beta$ deformation is larger in the triaxial than in the axial calculations. The largest differences
correspond, obviously, to the states that compose the $\gamma$ band in the triaxial case and to the $0^+_2$ state  due
to the fake minimum on the oblate side of Fig.~\ref{Fig1}.  Also interesting  to notice is that though the most probable $\gamma$-value  in the first band is zero, the average $\gamma$ values are around $15^\circ$. Also the average  $\gamma$ values for the $\gamma$-band are larger than the most probable ones. For completeness we also include the  values of the spectroscopic quadrupole moments.\\
At this point we would like to discuss the bands obtained in the RVAMPIR approach  and plotted in Fig.~\ref{Fig10}. At first glance both bands look similar to the corresponding ones of the full GCM calculations. A more careful analysis shows
that the GCM bands are slightly more compressed than the RVAMPIR ones in a better agreement with the experimental results. The 
transition probabilities are also very similar in both approaches.  It is really surprising that the RVAMPIR is able to provide
spectra and transition probabilities comparable to the full GCM approach.  There are several reasons which explain this behavior.
A look  to Fig.~\ref{Fig14} shows  that  all states of the first  and second band, respectively,  do have a similar probability distribution consisting of one maximum  -practically at the same ($\beta,\gamma$) point for all $I$-values-  and a homogeneous spread around this point. 
Such distributions can be very well approximated by a delta function at the given point. Furthermore since there is no $K$-mixing
neither in the GCM nor in the RVAMPIR there is no chance that the two approaches can differ in this respect.  The maxima of
band 2 are located practically at the same ($\beta,\gamma$) point  in both approaches  while for band 1 the GCM maximum
appears closer to the symmetry axis.  However since the surfaces are rather flat around these points this does not matter too much. 
 With respect to the $B(E2)$ transition probabilities we observe that they are rather similar to the ones of the GCM calculations, i.e.,
they are large for intraband and small for interband transitions. Interestingly the $2^{+}_{1}$, the $2^{+}_{2}$ and the $3^{+}_{1}$ RVAMPIR states  do have the same deformation parameters, see Table~\ref{table0}, i.e., we cannot argue, as in the GCM case, that the small interband transition probabilities are due to the poor overlap of the corresponding wave functions. The reason is 
 that the ground band is a pure $K=0$ and the $\gamma$-band a pure $|K|=2$ band. As a matter of fact, in this case, if we look at 
Eq.~\ref{trans_pro}, we observe  that if the factors sandwiched between the collective wave functions do not mix strongly the $K$
quantum number, then the transition probabilities are very small.  

Finally we compare the triaxial results with the available experimental data for $^{24}$Mg (see Fig. \ref{Fig12}). We find a remarkable qualitative agreement between theory and experiment both in energies and reduced transition probabilities. In both cases we observe a rotational ground state band, a second band associated to a $\gamma$-band and a third band with $\Delta I=2$. In fact, the theoretical description of the experimentally observed $\gamma$-band is one of the major achievements of the present model compared to previous implementations. Furthermore, in the particular case of $^{24}$Mg, the excitation energies within the ground  band are very well described even quantitatively with the present calculations as we see in Fig. \ref{Fig12}. In addition, it is important to emphasize the quality of the theoretical results for the intraband and interband reduced transition probabilities which reflect the small $K$ mixing between the corresponding bands. Although the improvement of the results with respect to the axial case is evident, the band heads of the  $\gamma$- and, especially, the third band are still calculated too high in excitation energy. This is probably due to the lack of the correlations associated to the angular momentum restoration before the variation and time-reversal symmetry breaking that are not included in this calculation. Additionally, the inclusion of two-quasiparticle states would further lower the excitation energies for these band heads.  These effects could also be present in the ground state bands. However, all these potential improvements are beyond the scope of the present work. They would  probably lead to a better quantitative description of the experimental results although we do not expect qualitative changes in the general picture.  Research  in this direction is in progress. \\
\begin{table}[htdp]
\centering
\begin{tabular}{cc|ccc|ccc}
\hline
\hline
$I$&$\sigma$&$\bar{\beta}_{tr}$&$\bar{\gamma}^{\circ}_{tr}$&$Q^{spec}_{tr}$&$\bar{\beta}_{ax}$&${\gamma}^{\circ}_{ax}$&$Q^{spec}_{ax}$\\
\hline
\hline
0&1&0.644 &14.81&0.00&0.510&0&0.00\\
0&2&0.515 &25.98&0.00&0.221&0&0.00\\
\hline
2&1&0.658 &13.58&-20.80&0.606&0&-19.94\\
2&2&0.709 &23.16&21.59&0.346&60&12.70\\
2&3&0.642 &16.02&-19.66&0.451&0&-19.33\\ 
\hline
3&1&0.717 &22.05&-0.18&-&-&-\\
\hline
4&1& 0.661&13.65&-26.33&0.623&0&-25.04\\ 
4&2& 0.714&22.60&-10.96&0.495&0&-20.33\\ 
4&3& 0.541&19.95&-18.51&-&-&-\\ 
\hline
5&1& 0.720&22.67&-17.51&-&-&-\\
\hline
6&1& 0.657&14.51&-28.42&0.622&0&-26.52\\
6&2& 0.702&22.92&-20.97 &-&-&-18.08\\
6&3&0.535 &17.24&-23.16&-&-&-\\ 
\hline
7&1& 0.730  &24.07&-24.49&-&-&-\\
\hline
8&1& 0.701 & 15.36&-25.96&0.648&0&-28.18\\
\hline
\hline
\end{tabular}
\caption{Average intrinsic deformation parameters and spectroscopic quadrupole moments (in e fm) in the triaxial and axial approximations.}
\label{table2}
\end{table}

\section{Summary}\label{summary}
In summary, we have presented the first implementation of the GCM-PNAMP method with fully triaxial intrinsic wave functions found by solving the VAP-PN equations with the Gogny interaction. Furthermore, due to the huge computational effort demanded by this type of calculations, we have established a protocol for a good performance of the method, namely:
\begin{enumerate}
\item Perform first a GCM-PNAMP with only axial $K=0$ wave functions in order to choose the number of oscillator shells, the relevant interval of  $\beta$ deformation and the density of points in the collective variable.
\item Study the convergence of the triaxial angular momentum projection with the number of integration points in the Euler angles by looking at the expectation value of $\hat{I}^{2}$ in the $(\beta,\gamma)$ plane and exploit the symmetries of the intrinsic states.
\item Choose a triangular mesh in the $(\beta,\gamma)$ plane in order to avoid both redundancy near the spherical shape and spurious effects due to a loss of resolution for increasing $\beta$.
\item  Select the converged states as the ones whose energy belongs to a \textit{plateau} and ensure the orthogonality with the other states with the same angular momentum.
\item Check the convergence of the results in the full triaxial calculation.
\end{enumerate} 
The method has been applied to the study of $^{24}$Mg which has been chosen as a test case in previous studies with different interactions. The comparison between axial and triaxial results shows minor changes in the ground state band which is predicted to be an axial rotational band with $K=0$. Only for angular momentum $I\geq4$ some $K$-mixing is observed giving rise to a small level compression. This result supports the use of axial calculations in these cases. However, the triaxial calculation is also able to reproduce the second band associated to a $\gamma$-band ($K=2$) observed  experimentally.  

 We have also introduced the RVAMPIR method which provides a more affordable alternative to the full GCM procedure for the calculation of ground and $\gamma$ bands. We  find that this approach provides a good description of the energy levels and the
intraband and interband  transition probabilities for the nucleus $^{24}$Mg. 

 Furthermore, the agreement between the theoretical and experimental results is in general good although some improvements beyond the scope of this work must be performed in order to give a better quantitative description. Some work is in progress in order to take into account these effects. In any case, $^{24}$Mg is not a very good example for studying strong triaxial effects like the ones mentioned in the introduction and the method will be applied in the near future to other systems where both triaxiality and $K$-mixing play a crucial role in describing the experimental data.
\begin{acknowledgments}
The authors acknowledge financial support from the Spanish Ministerio de Educaci\'on y Ciencia 
under contracts  FPA2007-66069 and FPA2009-13377-C02-01, by the Spanish Consolider-Ingenio 2010
Programme CPAN (CSD2007-00042) and within the Programa de Ayudas para Estancias de Movilidad Posdoctoral 2008 (T.R.R).
\end{acknowledgments}

\end{document}